\documentclass[11pt]{preprint}
\usepackage{times}
\usepackage{mhequ}
\usepackage{mhenvs}
\usepackage{ScriptFonts}
\usepackage{amssymb}
\usepackage{Symbols}
\usepackage{mhfig}
\let\eqref\eref
{\theorembodyfont{\rmfamily}
\newtheorem{step}{Step}
}
\newtheorem{example}[lemma]{Example}
\newtheorem{assumption}[lemma]{Assumption}
\def\eqdef{=:}

\newcommand{\cW}{{\cal W}}

\newcommand{\cA}{{\cal A}}

\newcommand{\cL}{{\cal L}}

\newcommand{\cC}{{\cal C}}
\newcommand{\cH}{{\cal H}}
\renewcommand{\epsilon}{\varepsilon}

\newcommand{\EX}{{\mathbb E}}
\newcommand{\E}{{\mathbb E}}

\def\Lip{{\mathrm{Lip}}}
\def\expect{{\EX}}

\begin{document}
\title{Modulation Equations:\\
Stochastic Bifurcation in Large Domains}
\author{D. Bl{\"o}mker\inst{1}, M. Hairer\inst{2}, and G. A. Pavliotis\inst{2}}
\institute{Institut f\"ur Mathematik, RWTH Aachen \and
Mathematics Research Centre, University of Warwick
 \\ \email{bloemker@instmath.rwth-aachen.de}\\
\email{hairer@maths.warwick.ac.uk} \\
\email{pavl@maths.warwick.ac.uk}}
 \date{\today}
\maketitle

\begin{abstract}
We consider the stochastic Swift-Hohenberg equation on a large domain
near its change of stability. We show that, under the appropriate scaling, its solutions
can be approximated by a periodic wave, which is modulated by the solutions
to a stochastic Ginzburg-Landau equation. We then proceed to show that this
approximation also extends to the invariant measures of these equations.
\end{abstract}

\tableofcontents

\section{Introduction}
We present a rigorous approximation result of stochastic partial differential equations
(SPDEs) by amplitude equations near a change of
stability. In order to keep notations at a bearable level, we focus on approximating
the stochastic Swift-Hohenberg equation by the stochastic Ginzburg-Landau equation,
although our results apply to a larger class of stochastic PDEs or systems of SPDEs.
Similar results are well-known in the deterministic case, see for instance
\cite{MR91j:35266,MR2001j:35030}.
However, there seems to be a lack of
theory when noise is introduced into the system. In particular, the treatment of
extended systems (\ie when the spatial variable takes values in an unbounded domain)
is still out of reach of current techniques.

In a series of recent articles
\cite{Bl-MP-Sc:01, DB-03, DB-Pquad, DB-Ha:PIM}, the  amplitude of the dominating pattern
was approximated by a stochastic ordinary differential equation (SODE).
On a formal level or without the presence of noise, the derivation of these results is well-known, see
for instance (4.31) or (5.11) in the comprehensive review article \cite{Cr-Ho:93} and references
therein.
This approach shows its limitations on large domains, where the spectral gap between the dominating
pattern and the rest of the equation becomes
small. It is in particular not appropriate to explain modulated pattern occurring in
many physical models and experiments (see \eg \cite{Ly:96, Ly-Mo:99}
or \cite{Cr-Ho:93} for a review). The  validity of the SODE-approximation is limited to a small
neighbourhood of the stability change, which shrinks, as the size of the domain gets
large.

For deterministic PDEs on unbounded domains it is well-known, see \eg
\cite{MR91j:35266,Mi-Sc:95,MR93i:35132,MR97a:35221},
that the dynamics of the slow modulations of the pattern can be described
by a PDE which turns out to be of Ginzburg-Landau type.

Since the theory of translational invariant SPDEs on unbounded domains
is still far from being fully developed, we adopt in the present article a
somewhat intermediate approach, considering large but bounded domains
in order to avoid the technical difficulties
arising for SPDEs on unbounded domains.
Note that the same approach has been used in \cite{MR2001j:35030} to study the
deterministic Swift-Hohenberg equation.
It does not seem possible to adapt the
deterministic theory directly to the stochastic equation.
One major obstacle is that the whole theory for
deterministic PDE relies heavily on good a-priori
bounds for the solutions of the amplitude equation
in spaces of sufficiently smooth functions.
Such bounds are unrealistic for our stochastic amplitude equation, since
it turns out to be driven by space-time white noise. Its solutions are therefore
only $\alpha$-H\"older continuous in space and time for $\alpha < 1/2$.
Nevertheless, the choice of large but bounded domains
captures and describes all the essential features of how noise in the
original equation enters the amplitude equation.

\subsection{Setting and results}
\label{sec:SR}

In this article, we concentrate on deriving the stochastic Ginzburg-Landau equation
as an amplitude equation for the stochastic Swift-Hohenberg equation. We will
discuss in Section~\ref{sec:similar} below for which class of equations similar
results are expected to hold. The Swift-Hohenberg equation is a
celebrated toy model for the convective instability in the Rayleigh-B\'enard
convection. A formal derivation of the equation from the
Boussinesq approximation of fluid dynamics can be found in \cite{Ho-Sw:77}.

In the following we consider solutions to
\begin{equ}[e:mainequ]\tag{SH}
\partial_t U= -(1+\d_x^2)^2 U+\epsilon^2 \nu U-U^3+\epsilon^{\frac32} \xi_\eps
\end{equ}
where  $U(x,t)\in \R$ satisfies periodic boundary conditions on $D_\epsilon =
[-L/\eps,L/\eps]$.
The noise $\xi_\eps$ is assumed to be real-valued homogeneous space-time
noise. To be more precise $\xi_\eps$ is a distribution-valued centred Gaussian field such
that
\begin{equ}[e:covorig]
\EX\xi_\eps(x,s)\xi_\eps(y,t)=\delta(t-s)q_\eps (|x-y|)\;.
\end{equ}
The family of correlation functions $q_\eps$ is assumed to converge in a suitable
sense to a correlation function $q$. One should think for the moment of $q_\eps$ as
simply being the $2L/\eps$-periodic continuation of the restriction of $q$ to $D_\eps$.
We will state in Assumption \ref{ass:gkeps}
the precise assumptions on $q$ and $q_\eps$. This will include
space-time white noise and noise with bounded correlation length.

The main result of this article is an approximation result for
solutions to \eref{e:mainequ}
by means of solutions to the stochastic Ginzburg-Landau equation.
We consider a class of ``admissible'' initial conditions given in Definition~\ref{def:admA}
below. This class is slightly larger than that of $\CH^1$-valued random variables with
bounded moments of all orders and is natural for the problem at hand.
In particular we show in \theo{thm:admu} the following.

\begin{theorem}{\bf (Attractivity)}\label{theo:attr}
Let $U$ be given by the solution of \eref{e:mainequ}
with arbitrary initial conditions, then there is a time $t_\eps>0$
such that for all $t\ge t_\eps$ the solution $U(t)$
is admissible.
\end{theorem}

Our main result (cf. Theorem \ref{thm:approx}) is the following:

\begin{theorem}\label{theo:main}{\bf (Approximation)}
Let $U$ be given by the solution of \eref{e:mainequ} with an admissible
initial condition written as $U_0(x) =
2 \epsilon \Re \bigl(a_0(\epsilon x) e^{i  x}\bigr)$. Consider the solution $a(X,T)$
to the stochastic Ginzburg-Lan\-dau equation
\begin{equ}[e:Amp]
\partial_T a= 4 \partial_X^2 a + \nu a -  3 |a|^2a + \sqrt{\hat q(1)}\, \eta\;,\quad
X\in[-L,L]\;,\; a(0) = a_0\;,
\end{equ}
where $\eta$ is complex space-time white noise and $\hat q$ denotes the Fourier
transform of $q$. Here, $a$ is subject to suitable boundary conditions, \ie those
boundary conditions such that $a(X,T) e^{i X/\eps}$ is $2L$-periodic.
Then, for every $T_0>0$, $\kappa > 0$, and $p\ge1$,
 one can find joint realisations of the noises $\eta$ and
$\xi_\eps$ such that
\begin{equ}[e:defapprox]
    \Big(\expect \sup_{\eps^2t \in [0,T_0]} \sup_{x \in D_\eps}
        |U(x,t) - 2 \epsilon \Re\bigl(a(\epsilon x, \epsilon^2 t) e^{i  x}\bigr)|^p\Big)^{1/p}
    \le C_{\kappa,p}\, \eps^{3/2 -\kappa}
\end{equ}
for every $\eps \in (0,1]$.
\end{theorem}
Note that solutions to \eref{e:mainequ} tend to be of order $\eps$, as can be seen from the fact that
this is the point where the dissipative nonlinearity starts to dominate the linear instability.
Therefore, the ratio between the size of the error and the size of the solutions is of order $\eps^{1/2}$.
Using an argument similar to the one in \cite{DB-Ha:PIM}, it is then straightforward
to obtain an approximation result on the invariant measures for \eref{e:mainequ} and
\eref{e:Amp}:
\begin{theorem}
\label{theo:IM1}
{\bf (Invariant Measures)}
Let $\nu_{\star,\eps}$ be the invariant measure for \eref{e:Amp} and let $\mu_{\star,\eps}$
be an invariant measure for \eref{e:mainequ}. Then, one can construct random
variables $a_\star$ and $U_\star$ with respective laws $\nu_{\star,\eps}$ and $\mu_{\star,\eps}$ such
that for every $\kappa > 0$ and  $p\ge1$
\begin{equ}
\Big(\expect \sup_{x \in D_\eps} |U_\star(x)
        - 2 \epsilon \Re \bigl(a_\star(\epsilon x) e^{ix}\bigr)|^p\Big)^{1/p}
    \le C_{\kappa,p}\, \eps^{3/2 - \kappa}\;,
\end{equ}
for every $\eps \in (0,1]$.
\end{theorem}
Let us remark that $\nu_{\star,\eps}$ is actually
independent of $\eps$, provided $L\in\eps\pi\N$.

Most of the present article is devoted to the proof of \theo{theo:main}.
We will then prove \theo{theo:attr} in Section \ref{sec:attr}
and \theo{theo:IM1} in Section~\ref{sec:inv_meas},
while  Section \ref{sec:stoch_conv} provides
a very general approximation result for linear equations,
that is used in the proof of \theo{theo:main}.

The remainder of this paper is organised as follows.
Section~\ref{sec:formal} is devoted to a formal justification of our results,
followed by a discussion on the type of equations for which
similar results are expected to hold. Note that the presented
approach relies on the presence of a stable cubic (or higher order)
nonlinearity.  For the moment, we cannot treat quadratic nonlinearities like
the one arising in the convection problems.
See however \cite{DB-Pquad} for a result on bounded domains covering that situation
or \cite{MR2000m:76061} for a deterministic result in unbounded domains.

The main step in the proof of \theo{theo:main} is then to define a
\textit{residual}, which measures how well a given process approximates
solutions to \eref{e:mainequ} via the variation of constants formula.
Section~\ref{sec:residual} provides estimates for this residual that are
used in Section~\ref{sec:approximation} to prove the main approximation result.

Section~\ref{sec:attr} justifies the assumptions on the initial conditions
required for the proof of
the approximation result, and Section~\ref{sec:inv_meas}
applies the result to the approximation of invariant measures.
The final Section~\ref{sec:stoch_conv} provides the approximation result
for linear equations in a fairly general setting.

\section{Formal Derivation of the Main Result}
\label{sec:formal}

In order to simplify notations, we work from now on with the rescaled version
$u(x,t)$ of the solutions of \eref{e:mainequ}, defined through $U(x,t) = \eps u(\eps
x, \eps^2 t)$. Then, $u$ satisfies the equation
\begin{equ}[e:meqres]
\partial_t u = -\eps^{-2}(1+\eps^2\d_x^2)^2 u+ \nu u - u^3 + \tilde \xi_\eps\;,
\end{equ}
with periodic boundary conditions on the domain $[-L,L]$. Here, we defined the
rescaled noise $\tilde \xi_\eps(x,t) = \eps^{-3/2}\xi_\eps(\eps^{-1}x, \eps^{-2}t)$.
This is obviously a real-valued Gaussian noise with covariance given by
$$
\EX\tilde \xi_\eps(x,t)\tilde \xi_\eps(y,s)
= \delta(t-s) \eps^{-1}q_\eps(\eps^{-1} |x-y|)\;.
$$
We define the operator $\CL_\eps = -1-\eps^{-2}(1+\eps^2\d_x^2)^2$
subject to periodic boundary conditions on $[-L,L]$ and we set $\tilde \nu = 1+\nu$, so that
\eref{e:meqres} can be rewritten as
\begin{equ}[e:orig2]\tag{SH$_\eps$}
\partial_t u = \CL_\eps u+ \tilde \nu u - u^3 + \tilde \xi_\eps\;.
\end{equ}
In oder to handle the fact that the dominating
modes $e^{\pm ix/\eps}$ are not necessarily $2L$-periodic,
we introduce the quantities
\begin{equ}
N_\eps = \Bigl[{ L \over \eps\pi}\Bigr]\;,
\quad
\delta_\eps = \frac1\eps -{\pi \over L}N_\eps\;,
\quad
\rho_\eps = N_\eps {\pi\eps \over L} \;,
\end{equ}
where $[\,x\,] \in \Z$ is used to denote the nearest integer of a real number
$x$ with the conventions that $[{1\over 2}] = {1 \over 2}$ and $[-x] = -[x]$.
Note that $|\delta_\eps|$ is therefore bounded by ${\pi\over 2L}$ independently of $\eps$.
Furthermore $\delta_\eps=0$ if and only if $L\in\eps\pi\N$.

With these notations, we rewrite the amplitude equation in a slightly different way.
Setting $A(x,t) = a(x,t)e^{i\delta_\eps x}$, \eref{e:Amp} is equivalent to
\begin{equ}[e:ampl2]\tag{GL}
\partial_t A= \Delta_\eps  A + \tilde \nu A -  3 |A|^2 A + \sqrt{\hat q(1)} \eta
\;,\quad
\Delta_\eps = - 1 - 4 (i\partial_x + \delta_\eps)^2\;,
\end{equ}
with \textit{periodic} boundary conditions,
where $\eta$ is another version of complex space-time white noise.
This transformation is purely only for
convenience, since periodic boundary conditions are more familiar.

Before we proceed further, we fix a few notations that will be used throughout
this paper. We will consider solutions to \eref{e:orig2} and \eref{e:ampl2} in
various function spaces, but let us for the moment consider them in $\L^2([-L,L])$.
We thus
denote by $\CH_u$ the $\L^2$-space of real-valued functions on $[-L,L]$ which
will contain the solutions to \eref{e:orig2}
and by
$\CH_a$ the $\L^2$-space of complex-valued functions on $[-L,L]$ which
will contain the solutions to \eref{e:ampl2}. In order to be
consistent with definitions \eref{e:defpi} and \eref{e:defiota} below, we define the
norm in $\CH_u$ as half of the
usual $\L^2$-norm, \ie
\begin{equ}
    \|u\|^2 = {1\over 2} \int_{-L}^L |u(x)|^2\,dx
        \;,\quad
    \|A\|^2 = \int_{-L}^L |A(x)|^2\,dx\;,
\end{equ}
for all $u \in \CH_u$ and all $A \in \CH_a$. We introduce the
projection $\pi_\eps:\CH_a \to \CH_u$ used in \eref{e:defapprox}, \ie
\begin{equ}[e:defpi]
    (\pi_\eps A)(x)
    = 2 \Re \bigl(A(x) e^{i\pi N_\eps x/L}\bigr)\;.
\end{equ}
We also define the injection $\iota_\eps:\CH_u \to \CH_a$ by
\begin{equ}[e:defiota]
(\iota_\eps u)(x) = u_+ \exp(-i\pi N_\eps x/L)\;,
\end{equ}
where, for $u = \sum_{k\in\Z} u_k \exp(i\pi k/L)$, we defined $u_+ =  \sum_{k>0}
u_k \exp(i\pi k/L) + {1\over 2} u_0$. Since $u$ is real-valued, one has of course
the equality $u = u_+ + \overline{ u_+ }$,
where $\overline{u_+}$ denotes the complex conjugate of $u_+$. Furthermore, one has the relations
\begin{equ}[e:identity]
\pi_\eps \circ \iota_\eps = \iota_\eps^* \circ \iota_\eps^{} = \text{Id}\;,
\end{equ}
and the embedding $\iota_\eps$ is isometric. Here, $\iota_\eps^*:\CH_u \to \CH_a$ denotes
the adjoint of $\iota_\eps$. We also define the space $\CH_{\iota} \subset \CH_a$ as the
image of $\iota_\eps$. Equation \eref{e:identity} implies in particular that
$\pi_\eps = \iota_\eps^*$, if both operators are restricted to $\CH_{\iota}$.
Note also that $\iota_\eps$
is \textit{not} a bounded operator between the corresponding $\L^\infty$ spaces,
even though $\pi_\eps$ is.

With these notations in mind, we give a formal argument that shows why \eref{e:ampl2} is
expected to yield a good approximation for \eref{e:orig2}. First of all, note that
even though $\iota_\eps \circ \pi_\eps$ is not the identity, it is close to the
identity when applied to a function which is such that its Fourier modes with
wavenumber larger than $\eps^{-1}$ are small.
This is indeed expected to be the case for the solutions $A$ to
\eref{e:ampl2}, since the heat semigroup $e^{\Delta_\eps t}$ strongly damps high frequencies.

\begin{figure}\label{fig:spectr}
\begin{center}
\mhpastefig[9/10]{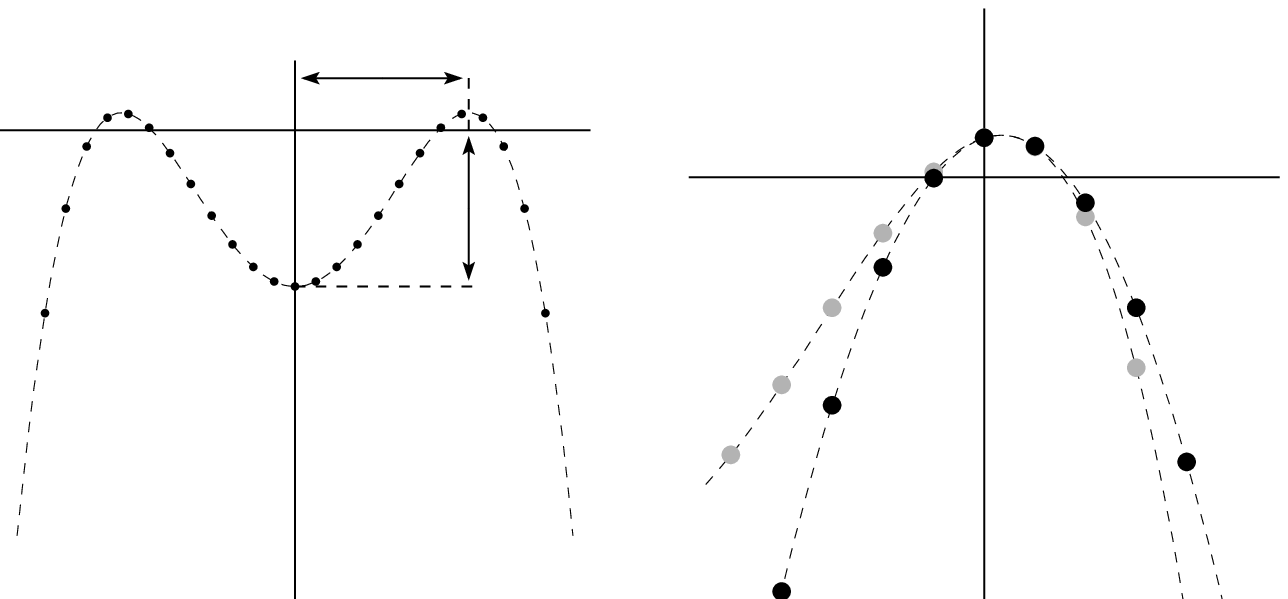}
\end{center}
\caption{Spectra of the linear parts.}
\end{figure}

Hence, $\iota_\eps \pi_\eps A\approx A$.
Therefore, making the ansatz $u = \pi_\eps A$ and plugging it into
\eref{e:orig2} yields
\begin{equ}
\d_t A \approx \iota_\eps \CL_\eps \pi_\eps A + \tilde\nu A
- \iota_\eps (\pi_\eps A)^3 + \iota_\eps \tilde \xi_\eps\;.
\end{equ}
The left part of Figure~\ref{fig:spectr} shows the spectrum of $\tilde \nu +
\CL_\eps$. The right part shows the spectrum of $\iota_\eps(\tilde \nu +
\CL_\eps)\pi_\eps$ (which is interpreted as a self-adjoint operator from
$\CH_{\iota}$ to $\CH_{\iota}$)
in grey and the spectrum of $\Delta_\eps + \tilde \nu$ in black.
One sees that the two are becoming increasingly similar as $\eps \to 0$, since the
tip of the curve becomes increasingly well approximated by a parabola.

Expanding the term $(\pi_\eps A)^3$ we get
\begin{equ}
(\pi_\eps A)^3
    = A^3 e^{{3i\pi N_\eps}x/L}
    + 3 A |A|^2e^{{i\pi N_\eps}x/L}
    +   3\bar A |A|^2e^{-{i\pi N_\eps}x/L} +
\bar A^3 e^{-{3i\pi N_\eps}x/L}\;.
\end{equ}
Therefore, one has
\begin{equ}
\iota_\eps (\pi_\eps A)^3 \approx A^3 e^{{3i\pi N_\eps}x/L} + 3 A |A|^2\;.
\end{equ}
Since the term with high wavenumbers will be suppressed by the linear part, we can
arguably approximate this by $3 A|A|^2$, so that we have
\begin{equ}[e:myapprox]
    \d_t A \approx
    \Delta_\eps  A + \tilde \nu A -  3 |A|^2 A + \iota_\eps \tilde \xi_\eps\;.
\end{equ}
It remains to analyse the behaviour of $\iota_\eps \tilde \xi_\eps$ in the limit of small
values of $\eps$. Note that we can expand $\tilde \xi_\eps$ in Fourier series, so that
\begin{equ}
    \tilde \xi_\eps(x,t)
    \stackrel{\mbox{\tiny law}}{=}
    c_L\sum_{k\in\Z} \sqrt{\hat q_\eps (\eps k\pi/L)}  \xi_k(t) e^{i k \pi x/L}\;,
\end{equ}
where the $\xi_k(t)$ denote complex independent white noises,
with the restriction that $\xi_{-k}=\overline{\xi_k}$, and where we set $c_L = 1/\sqrt{2L}$.
On a formal level, this yields
for $\iota_\eps \tilde \xi_\eps$
\begin{equs}
\iota_\eps \tilde \xi_\eps(x,t)
&\stackrel{\mbox{\tiny law}}{\approx}  \sum_{k=0}^\infty c_L\sqrt{\hat q(\eps k\pi/L)}
     \xi_k(t) e^{i\pi(k -N_\eps) x/L}\\
&\stackrel{\mbox{\tiny law}}{=}
    c_L \sum_{k= - N_\eps}^\infty \sqrt{\hat q(\pi\eps(N_\eps+k)/L)} \xi_k(t)e^{i\pi k x/L} \\
 &\approx c_L \sum_{k \in \Z}^\infty \sqrt{\hat q(1)} \xi_k(t)  e^{i \pi k x/L}
    \approx \sqrt{\hat q(1)}\, \eta(x,t)\;.
\end{equs}
In this equation, we justify the passage from the second to the
third line by the fact that the linear part of \eref{e:ampl2} damps high
frequencies, so contributions from Fourier modes beyond $k \approx \eps^{-1}$
can be neglected. Furthermore, $\pi\eps(N_\eps+k)/L\to1$ for $\eps\to0$.

Plugging the previous equation into \eref{e:myapprox}, we obtain \eref{e:ampl2}.
The aim of the present article is to make this formal
calculation rigorous.

\subsection{Extension of our results}
\label{sec:similar}

Even though we restrict ourselves to the case of the stochastic Swift-Hohenberg
equation, it is clear from the above formal calculation that one expects similar
results to hold for a much wider class of equations. For example the linear result
holds for a quite general class of operators (cf. Section \ref{sec:stoch_conv}).
Consider a smooth even function
$P:\RÊ\to \R_+$ which grows sufficiently fast
at infinity and has a finite number of zeroes. Consider
also a stable (\ie a dissipativity condition of the type $\scal{u, \CF(u)} < 0$ holds)
cubic nonlinearity $\CF$ of the type
\begin{equ}
\CF(u) = g \star (g_1 \star u) (g_2 \star u) (g_3 \star u)\;,
\end{equ}
for some distributions $g, g_i$, where $\star$ denotes the convolution. One could
for example choose $\CF(u) = -u |\nabla u|^2$, $\CF(u) = \Delta u^3$,
or $\CF(u) = \nabla |\nabla u|^3$.

Provided that
the Fourier transforms of the $g$'s are smooth and of sufficiently slow growth
relative to the growth of $P$, the equation
\begin{equ}[e:mainequgen]
\partial_t U= -P(i \d_x)\, U+\epsilon^2 \nu U - \CF(U)+\epsilon^{\frac32} \xi_\eps\;,
\end{equ}
with periodic boundary conditions on $D_\epsilon =
[-L\epsilon^{-1},L\epsilon^{-1}]$, has a unique global solution. This solution can
then be approximated in the same fashion by
\begin{equ}[e:amplequgen1]
U(x,t) = 2\epsilon \sum_{i=1}^n \Re\bigl(A_i(\epsilon x, \epsilon^2 t) e^{i \nu_i
x}\bigr) + \CO(\eps^{3/2})\;,
\end{equ}
where the values $\nu_i$ correspond to the zeroes of $P$ and where the $A_i$ solve a
finite number of coupled stochastic Ginzburg-Landau equations.
Let us illustrate this by two examples that cover the most typical situations. Take
\begin{equ}[e:example]
\partial_t U= -(\d_x^2 + 1)^2(\d_x^2 + 9)^2 U+\epsilon^2 U - U |\d_x U|^2 + \epsilon^{\frac32} \xi_\eps\;.
\end{equ}
Then, one has
\begin{equ}[e:amplequgen2]
U(x,t) =
  2\epsilon \Re \bigl(A_1(\epsilon x, \epsilon^2 t) e^{i x}\bigr)
+ 2\epsilon\Re \bigl(A_2(\epsilon x, \epsilon^2 t) e^{3 i x}\bigr)
+ \CO(\eps^{3/2})\;,
\end{equ}
where the amplitudes $A_1$ and $A_2$ satisfy
\begin{equs}
\d_t A_1 &= \alpha_1 \Delta A_1 + A_1 - 3 A_1 |A_1|^2 - 18 A_1 |A_2|^2 + 5 A_2 \bar A_1^2 + \sqrt{\hat q(1)}\,\eta_1 \\
\d_t A_2 &= \alpha_2 \Delta A_2 + A_2 -2 A_2 |A_1|^2 - 27 A_2 |A_2|^2 - A_1^3 +
\sqrt{\hat q(3)}\,\eta_2\;,
\end{equs}
for some coefficients $\alpha_i$. Here, the $\eta_i$ are two independent space-time
white noises. The term $A_1^3$ is a ``resonance'' that comes from the fact that
$(e^{ix})^3 = e^{3ix}$. The term $A_2 \bar A_1^2$ similarly comes from the fact that
$e^{3ix} (e^{-ix})^2 = e^{ix}$.
If \eref{e:example} is replaced \eg by
\begin{equ}
\partial_t U= -(\d_x^2 + 1)^2(\d_x^2 + 16)^2 U+\epsilon^2 U - U |\d_x U|^2 + \epsilon^{\frac32} \xi_\eps\;,
\end{equ}
then these resonances disappear and one has
\begin{equ}[e:amplequgen3]
U(x,t) = \epsilon \Re \bigl(A_1(\epsilon x, \epsilon^2 t) e^{i x}\bigr) + \epsilon
\Re \bigl(A_2(\epsilon x, \epsilon^2 t) e^{4 i x}\bigr) + \CO(\eps^{3/2})\;,
\end{equ}
where the amplitudes $A_1$ and $A_2$ satisfy
\begin{equs}
\d_t A_1 &= \tilde\alpha_1 \Delta A_1 + A_1 - 3 A_1 |A_1|^2 - 18 A_1 |A_2|^2 + \sqrt{\hat q(1)}\,\eta_1 \\
\d_t A_2 &= \tilde\alpha_1 \Delta A_2 + A_2 -2 A_2 |A_1|^2 - 27 A_2 |A_2|^2 +
\sqrt{\hat q(4)}\,\eta_2\;,
\end{equs}
for some coefficients $\tilde \alpha_i$.
Before we proceed with the proofs of the results stated in the introduction, let us introduce a
few more notations that will be useful for the rest of this article.

\subsection{Notations, projections, and spaces}

We already introduced the $\L^2$-spaces $\CH_a$ and $\CH_u$, as well as the operators
$\pi_\eps$ and $\iota_\eps$.
We will denote by $e_k(x) = e^{i k\pi x/L}/\sqrt{2L}$
the complex orthonormal Fourier basis in
$\CH_a$.
\begin{definition}
\label{def:sobolev}
We define
the scale of (fractional) Sobolev spaces $\CH_a^\alpha \subset \CH_a$ with $\alpha \in \R$
as the closure of the set of
$2L$-periodic complex-valued trigonometric polynomials $A = \sum A_k e_k$
under the norm  $\|A\|_\alpha^2 = \sum_k (1+|k|)^{2\alpha} |A_k|^2$.
We also define the space $\CH_u^\alpha$ as those real-valued functions $u$ such that
$\iota_\eps u \in \CH_a^\alpha$. We endow these spaces with the natural norm
$\|u\|_\alpha = \|\iota_\eps u\|_\alpha$.
\end{definition}
We also denote by $\L^p_a$ (respectively $\L^p_u$) the complex (respectively real) space $\L^p([-L,L])$,
endowed with the usual norm. We similarly define the spaces $\CC^0_a$ and $\CC^0_u$ of
periodic continuous bounded functions.
We will from time to time consider $e_k$ as elements of $\CH_a^\alpha$, $\L^p_a$,
or the complexifications of $\CH_u^\alpha$ and $\L^p_u$.

Note that with this notation, we have
\begin{equ}
\iota_\eps \pi_\eps e_k = \left\{\begin{array}{ll} e_k & \text{if $k \ge -N_\eps$,} \\ e_{-k-2N_\eps} & \text{if $k < -N_\eps$.} \end{array}\right.
\end{equ}
In particular, one has $\|\pi_\eps e_k\|_\alpha \le \|e_k\|_\alpha$ for every $\alpha \ge 0$.

\begin{remark}
Although the norm in $\CH_u^\alpha$ is equivalent to the standard $\alpha$-Sobolev norm,
the equivalence constants depend on $\eps$.
In particular, the operators $\iota_\eps:\CH_u^\alpha\to \CH_a^\alpha$
and $\pi_\eps:\CH_a^\alpha\to\CH_u^\alpha$
are bounded by $1$ with our choice of norms,
which would not be the case if $\CH_u^\alpha$ was equipped with the standard norm instead.
\end{remark}

\begin{remark}
\label{rem:H1intoC0}
Since the injection $\iota_\eps:\CH_u^1 \to \CH_a^1$, the inclusion $\CH_a^1 \hookrightarrow \CC^0_a$, as well as the projection $\pi_\eps : \CC^0_a \to \CC^0_u$ are all bounded independently of $\eps$, the inclusion $\CH_u^1 \hookrightarrow \CC^0_u$, which is given by the composition of these three operators, is also bounded independently of $\eps$.
\end{remark}
Finally, we define, for some sufficiently small constant $\delta>0$, the projections
$\Pi_{\delta/\eps}$ and  $\Pi^c_{\delta/\eps}$by
\begin{equation}
\Pi_{\delta/\eps}\Big( \sum_{k\in\Z}\gamma_k e^{ik\pi x/L}\Big)=
\sum_{|k|\le \delta/\eps} \gamma_k e^{ik\pi x/L}
\quad\mathrm{and} \quad
\Pi^c_{\delta/\eps}=1-\Pi_{\delta/\eps}\;.
\label{e:proj}
\end{equation}
%
%
%
\section{Bounds on the Residual}
\label{sec:residual}
Our first step in the proof of \theo{theo:main} is to control the
residual (defined in Definition~\ref{def:residual} below), which measures how well a given
approximation satisfies the mild formulation of \eqref{e:orig2}. Before we give the
definition of a mild solution, we define
the stochastic convolutions $ W_{\cL_\eps}(t)$ and $ W_{\Delta_\eps}(t)$, which are
formally the solutions to the linear equations:
\minilab{e:stochconv}
\begin{equs}
W_{\cL_\eps}(t) &=  \sqrt{Q_\eps} \int_0^t e^{(t - \tau) \cL_\eps} \, d W_\xi (t)
\label{e:stochconv1}\\ \quad
W_{\Delta_\eps}(t) &= \sqrt{\hat{q}(1)}\int_0^t e^{(t - \tau) \Delta_\eps} \, d W_\eta (t)\;.
\label{e:stochconv2}
\end{equs}
Here $W_\xi(t)$ and $W_\eta(t)$ denote standard cylindrical Wiener
processes (\ie space-time white noises). Note that $W_\xi$ is real valued,
while $W_\eta$ is complex valued.

The definition of the covariance operator $Q_\eps$ is given in
Definition \ref{def:cov_op} and is such that $\sqrt{Q_\eps} \partial_t W_\xi$ has the covariance
structure given in \eref{e:covorig}.
We will assume throughout the paper
that Assumption \ref{ass:gkeps} holds for the correlation functions $q$ and $q_\eps$.
In particular, note that $Q_\eps$ is a convolution operator and therefore commutes with
the semigroup generated by $\cL_\eps$.

With these notations, a mild solution, see \eg \cite[p. 182 ]{ZDP1}, of the rescaled equation
\eref{e:orig2} is a process $u$ with continuous paths such that:
\begin{equation}\label{e:uvoc}
   u(t)= e^{t\cL_\eps}u(0)+\int_0^t e^{(t-\tau)\cL_\eps}\bigl(\tilde\nu
u(\tau)-u^3(\tau)\bigr) d\tau  + W_{\cL_\eps}(t)\;,
\end{equation}
almost surely. We also consider mild solutions $A$ of \eref{e:ampl2}
\begin{equation}\label{e:Avoc}
   A(t)= e^{t\Delta_\eps}A(0)+\int_0^t e^{(t-\tau)\Delta_\eps}\bigl(\tilde\nu
A(\tau)-3 |A(\tau)|^2 A(\tau) \bigr) \, d\tau  + W_{\Delta_\eps}(t)\;.
\end{equation}
This motivates the following definition:

\begin{definition}\label{def:residual}
Let $\psi$ be an $\cH_u$-valued process. The
\textit{residual} $\mathrm{Res}(\psi)$ of $\psi$ is the process given by
\begin{equation}
    \mathrm{Res}(\psi)(t)=
    -\psi(t)+e^{t\cL_\eps}\psi(0)+\int_0^t e^{(t-\tau)\cL_\eps}\bigl(\tilde\nu
\psi(\tau)-\psi^3(\tau)\bigr) d\tau  + W_{\cL_\eps}(t)\;,
\end{equation}
where $W_{\cL_\eps}(t)$ is as in \eref{e:stochconv1}.
\end{definition}

It measures how well the process $\psi$ approximates a mild solution of \eqref{e:orig2}.
Let us now introduce the concept of admissible initial condition. Since we are
dealing with a family of equations parametrised by $\eps \in (0,1)$, we actually
consider a family of initial conditions. We emphasise on the $\eps$-dependence
here, but we will always consider it as implicit in the sequel.
\begin{definition}
\label{def:admA}
A family of random variables $A^\eps$ with values in $\CH_a$ (or equivalently a
family $\mu^\eps$ of probability measures on $\CH_a$)
is called \textit{admissible} if there exists a decomposition $A^\eps = W_0^\eps + A_1^\eps$,
a constant $C_0$, and a family of constants $\{C_q\}_{q\ge1}$ such that
\begin{claim}
\item[1.] $A_1^\eps \in \CH_a^1$ almost surely and $\EX \|A_1^\eps\|_{1}^{q} \le C_q$ for every $q \ge 1$,
\item[2.] the $W_0^\eps$ are centred Gaussian random variables such that
\begin{equation}
\bigl| \EX \langle e_k, W_0^\eps \rangle \langle e_{\ell}, W_0^\eps \rangle\bigr| \leq
C_0 \frac{\delta_{k \ell}}{1 + |k|^2} \;,
\label{e:noise_adm}
\end{equation}
for all $k,\ell \in \Z$,  ($\delta_{k \ell}=1$ for $k=\ell$ and $0$ otherwise)
\end{claim}
and such that these bounds are independent of $\eps$. A family of random variables
$u^\eps$ with values in $\CH_u$ is called admissible if $\iota_\eps u^\eps$ is admissible.
\end{definition}

\begin{remark}
The definition above is consistent with the definition of $\pi_\eps$ in the sense that if
$A^\eps$ is admissible, then $\pi_\eps A^\eps$ is also admissible.
\end{remark}

\begin{remark}\label{rem:noise}
Note that \eqref{e:noise_adm} implies that the covariance operator of $W_0^\eps$ commutes with
the Laplacian, so that $W_0^\eps \stackrel{\mbox{\tiny law}}{=} \sum_{k\in \Z} c_k^\eps \xi_k e_k$,
where $c_k^\eps \le C/(1+|k|)$ and
the $\xi_k$ are independent normal random variables
with the restriction that $\xi_{-k}=\overline{\xi_k}$.
This implies by \lem{lem:Dirk} that $\expect \|W_0^\eps\|_{\CC_a^0}^p\le C$ for every $p\ge 1$,
as $\|e_k\|_{L^\infty}\le C$ and $\Lip(e_k)\le C|k|$.
\end{remark}
We have the following result.
\begin{theorem} {\bf (Residual)}
\label{thm:res} For every $p \ge 1$, $T_0>0$, $\kappa > 0$, and admissible initial
condition $A(0)$, there is a constant $C_{\kappa, p}>0$ such that the mild solution $A$
of \eqref{e:ampl2} with initial condition $A(0)$ satisfies
\begin{equation}
 \EX \Big( \sup_{t \in[0,T_0 ]}\|\mathrm{Res}(\pi_\eps A)(t)\|_{\CC_u^0}^p\Big)
 \le C_{\kappa,p}\, \epsilon^{\frac{p}{2} - \kappa}.
\label{e:est_resid}
\end{equation}
\end{theorem}
For the proof of the theorem we need two technical lemmas. The first one provides
us with estimates on the operator norm for the difference between the semigroup of the original
equation and that of the amplitude equation.
\begin{lemma}
\label{lem:sem_dif}
Let $H_t$ be defined as
\begin{equation}
H_t := e^{- \cL_\eps t} \pi_{\eps} - \pi_{\eps} e^{- \Delta_\eps t}\;.
\label{e:sem_dif}
\end{equation}
Then for all  $\alpha>0$ there exists a constant $C >0$ such that
\begin{equ}
\|H_t \|_{\cL(\cH_a, \cH^{\alpha}_u)} \leq C \eps t^{-\frac{\alpha +1}{2}}\quad\text{and}\quad
\|H_t \|_{\cL(\cH_a^1, \CC^{0}_u)} \leq C \eps^{1/2}\;.
\label{e:op_norm}
\end{equ}
\end{lemma}
\begin{proof}
The operator $H_t$ acts on $e_k\in\CH_a$
as
\begin{equation}
H_t e_k = \lambda_k(t)\,\pi_\eps e_k\;,
\label{e:Htonek}
\end{equation}
 where
the $\lambda_k(t)$'s are given by
\begin{equ}[e:lambdak]
\lambda_k(t)
= c e^{-t \left(1+\eps^{-2} \left(1-\frac{\eps^2 \pi^2}{L^2}(k-N_\eps)^2 \right)^2 \right)}
- c e^{-t \left(1+4 \left(\frac{ k\pi}{L} -\delta_\eps \right)^2 \right)},
\end{equ}
with some constant $c$ bounded by $1$.
By Taylor expansion around $k=0$,
we easily derive for some constants $c$ and $C$ the bound
\begin{equ}[e:boundlambda]
|\lambda_k(t)|
\le \left\{\begin{array}{cl}
C & \text{for all $k\in\Z$,} \\
C t \eps (1+|k|)^3 e^{- c t (1+|k|)^2}& \text{for $|k| \le N_\eps$,}
\end{array}\right.
\end{equ}
Let now $h = \sum_{k \in \Z} h_k e_k \in \cH_a$.
We write
$\|H_t h \|_{\alpha} \le \|H_t \Pi_{\delta/\eps} h \|_{\alpha}
+ \|H_t \Pi_{\delta/\eps}^c h \|_{\alpha}$
for $\delta>0$ sufficiently small such that $\delta/\eps \le N_\eps$.
It follows furthermore from standard analytic semigroup theory
that $H_t$ is bounded by $C t^{-(\alpha+1)/2}$ as an operator from $\CH_a^{-1}$
into $\CH_u^\alpha$.
Since the operator $\Pi_{\delta/\eps}^c:\CH_a \to \CH_a^{-1}$ is bounded by $C\eps$,
it follows that one has indeed
$\|H_t \Pi_{\delta/\eps}^c h \|_{\alpha} \le C\eps t^{-(\alpha + 1)/2}\|h\|
$.
The term $\|H_t \Pi_{\delta/\eps} h \|_{\alpha}$ is in turn bounded by
\begin{equs}
\|H_t  \Pi_{\delta/\eps} h \|_{\alpha}^2 & \leq
C t^2 \eps^2 \sum_{|k| \le \delta/\eps} (1+|k|)^{6 + 2 \alpha} e^{-c t (1+|k|)^2} |h_k |^2
 \\ & \leq
C t^{-\alpha -1} \eps^2 \sum_{|k| \le \delta/\eps} \bigl(t(1+|k|)^2 \bigr)^{3 + \alpha} e^{-c t (1+|k|)^2} |h_k |^2 \\
    & \leq C t^{-\alpha -1} \eps^2 \|h\|^2\;,
\end{equs}
from which the first bound follows. To show the second bound, take
$h = \sum_{k} h_k e_k$ in $\cH_a^1$.
Now a crude estimate shows
\begin{equ}[e:Htf]
\|H_t h \|_{\CC_u^0}
\le C\sum_{k \in \mathbb{Z}} |\lambda_k(t)|\,|h_k|
\le C\sqrt{\sum_{k \in \mathbb{Z}} {|\lambda_k(t)|^2\over 1+|k|^2}} \|h\|_1\;.
\end{equ}
It follows from \eref{e:boundlambda} that
\begin{equ}[e:bouLk]
{|\lambda_k(t)|^2 /(1+|k|^2)}
\le C \min\{\eps^2 , 1/(1+|k|^2)\}\;,
\end{equ}
so that
$\sum_{k \in \mathbb{Z}} {|\lambda_k(t)|^2\over 1+|k|^2} \le C \eps$
by treating
separately the case $|k|\le \eps^{-1}$ and the case $|k|> \eps^{-1}$.
\end{proof}
The second technical lemma bounds the difference between the linear part of the original
equation and that of the amplitude equation, applied to an admissible initial condition.
The idea is that, for an initial condition which admits the decomposition
$A = W_0 + A_1$, one can use the $\CH_a^1$-regularity to control the term involving $A_1$
and Gaussianity to control the term involving $W_0$.
\begin{lemma} \label{lem:ic}
Let $A$ be admissible in the sense of Definition \ref{def:admA}
and let $H_t$ be defined by \eqref{e:sem_dif}.
Then for every $T_0 >0$ $\kappa>0$ and $p \ge 1$ there exist constants $C>0$ such that
\begin{equ}[e:est_ic]
    \E\Big(\sup_{t \in [0,T_0]} \| H_t A\|_{\CC_u^0}^p\Big)
    \leq C \eps^{\frac{p}{2} - \kappa}.
\end{equ}
\end{lemma}
\begin{proof}
Since $A$ is admissible, it can be written as $A= W_0 + A_1$ with the same notations
as in Definition~\ref{def:admA}. The bound on $H_t A_1$ is an immediate consequence of \lem{lem:sem_dif}
above, so we only consider the term involving $W_0$. We write $W_0 =  \sum_{k\in \Z} c_k^\eps \xi_k e_k$ as in Remark~\ref{rem:noise}, so that by \eref{e:Htonek}
\begin{equ}
H_t W_0 = \sum_{k\in \Z} c_k^\eps \lambda_k(t) \xi_k \,\pi_\eps e_k\;,
\end{equ}
with $\lambda_k$ as in \eref{e:lambdak}.
We use now Lemma \ref{lem:Dirk} with domain $G = [-L,L] \times [0,T_0]$ and
\begin{equ}
f_k(x,t) = c_k^\eps \lambda_k(t) \, (\pi_\eps e_k)(x)\;.
\end{equ}
From \eref{e:bouLk}, we derive $\|f_k\|_{\L^\infty} \le C \min\{\eps , 1/(1+|k|)\}$.
 Furthermore, it is
easy to see by a crude estimate on $\Lip(\lambda_k)$
that $\Lip(f_k) \le C\eps^{-4} (1+|k|)^4$ for some constant $C$, so
that the required bound follows. Note that any bound on $\Lip(f_k)$  which is polynomial
in $\eps^{-1}$ and $|k|$ is sufficient.
\end{proof}
\begin{proof}[of Theorem \ref{thm:res}]
We start be reformulating the residual in a more convenient way. We add and subtract
$\int_0^t e^{(t-\tau) \cL_\eps} (\pi_\eps 3A|A|^2)(\tau) \, d \tau$ to obtain
\begin{equs}
\mathrm{Res}(\pi_\eps A)(t)  &= -(\pi_\eps A)(t) +  e^{t\cL_\eps} (\pi_\eps A)(0) + W_{\cL_\eps}(t)
 \nonumber   \\ &
\quad + \tilde\nu  \int_0^t e^{(t-\tau)\cL_\eps} \left(\tilde\nu (\pi_\eps A)(\tau) - ((\pi_\eps A)(\tau))^3
\right) d\tau
\nonumber         \\  & =  H_t A(0) + \int_0^t H_{t - \tau} \left( \tilde\nu\eps A(\tau) -
     ( A(\tau))^3 \right) d\tau  \nonumber \\ &
\quad + \int_0^t e^{(t-s)\cL_{\epsilon}} \left( ( \pi_\eps 3|A|^2 A)(\tau) -((\pi_\eps A)(\tau))^3 \right)
d \tau  \nonumber \\ & \quad +  W_{\cL_\eps}(t) -  \pi_\eps W_{\Delta_\eps}(t) \, ,
\nonumber
 \end{equs}
where the operator $H_t$ is defined in \eqref{e:sem_dif}. We estimate each term in the above expression
separately, starting with the one involving the initial conditions.
Since we have assumed that $A(0)$ is admissible, Lemma
\ref{lem:ic} applies and we obtain
\begin{equ}
\E\sup_{t \in [0,T]} \| H_t A(0)\|_{\CC^0_u}^p
 \leq C_p \eps^{\frac{p}{2}- \kappa}.
\end{equ}
Furthermore, Proposition \ref{prop:stoch_conv} shows that there is a way of correlating the stochastic convolutions such that:%
\begin{equ}
\E\sup_{t \in [0,T]} \|  W_{\cL_\eps}(t) -  \pi_\eps W_{\Delta_\eps}(t)\|_{\CC^0_u}^p
\leq C_p \eps^{\frac{p}{2}- \kappa}.
\end{equ}
We  now use Lemma \ref{lem:sem_dif} for some $\alpha\in(\frac12,1)$
together with the embedding of $\CH_a^\alpha$ in $\CC^0_a$  to deduce that:
\begin{equs}
\Bigl\| \int_0^t H_{t - \tau}  \Bigl( \tilde \nu \eps A(\tau) -
     ( A(\tau))^3 \Bigr) d\tau  \Bigr\|_{\cC^0_u}
    & \leq C \int_0^t \|H_{t - \tau} \|_{\cL(L^2_a,\cH^{\alpha}_a)} \, d \tau
    \sup_{0 \leq \tau \leq t} \|A(\tau) \|_{\L^6_a}^3
 \\ & \leq
        C \eps \int_0^t (t - \tau)^{- \frac{\alpha +1}{2}} \, d \tau
        \sup_{0 \leq \tau \leq t} \|A(\tau) \|_{\L^6_a}^3\\
     &\leq C \eps \sup_{0 \leq \tau \leq t} \|A(\tau) \|_{\L^6_a}^3\;.
\end{equs}
Thus with the a--priori estimate on the solution of the amplitude equation from
Proposition \ref{prop:apriori}
\begin{equ}
\E\sup_{t \in [0,T]} \Bigl\| \int_0^t H_{t - \tau}  \left( (\nu +1) \eps A(\tau) -
     ( A(\tau))^3 \right) d\tau  \Bigr\|_{\CC^0_u}^p \leq C_p \eps^{p}\;.
\end{equ}
Let us turn to the remaining term.
We have (writing $\tilde e_{2N_\eps} = e^{{2 i \pi N_\eps x/L}} $)
\begin{equs}
\int_0^t e^{(t-\tau)\cL_{\epsilon}} \bigl( 3 \pi_\eps \bigl(|A|^2 A\bigr)(\tau) &-  \bigl(\pi_\eps A(\tau)\bigr)^3 \bigr)
d \tau    =   \int_0^t e^{(t-\tau)\cL_{\epsilon}} \pi_\eps \left( A(\tau)^3 \tilde e_{2N_\eps}
\right) d \tau  \\
& =
\int_0^t \pi_\eps e^{(t-\tau)\Delta_{\eps}} \left( A(\tau)^3 \tilde e_{2N_\eps}
\right) d \tau \\
&\quad+ \int_0^t H_{t - \tau} \left( A(\tau)^3 \tilde e_{2N_\eps}
\right) d \tau. \nonumber \\ & =:  I_1(t) + I_2(t). \nonumber
\end{equs}
Let us consider first $I_2(t)$. We use Lemma \ref{lem:sem_dif},
together with the \textit{a priori} estimate on $A$ from
Proposition \ref{prop:apriori} to obtain:
\begin{equ}
\E\sup_{t \in [0,T]} \|  I_2(t)\|_{\cC_u^0}^p \leq C_p \eps^{p}.
\end{equ}
Now we turn to $I_1(t)$. By Theorem \ref{thm:fourier}, since we have assumed
that the initial conditions are admissible,
we know  that $A(t)$ is concentrated in Fourier space:
$$
\EX  \sup_{t \in[0,T_0]} \|\Pi^c_{\delta/\eps}A(t) \|_{\cC_a^0}^p \le C \eps^{\frac{p}{2} -
\kappa}.
$$
Consequently we have $A^3=(\Pi_{\delta/\eps}A)^3 + Z$, where
\begin{equ}[e:lastone]
\EX  \sup_{t \in[0,T_0]} \|Z \|_{\cC_a^0}^p \le C \eps^{\frac{p}{2} -
\kappa} \quad \mbox{and} \quad
\EX  \sup_{t \in[0,T_0]} \|\Pi_{\delta/\eps}A(t) \|_{\cC_a^0}^p \le C.
\end{equ}
Furthermore, we know that $(\Pi_{\delta/\eps}A)^3 e_{2N_\eps}$ has non-vanishing Fourier
coefficients only for wavenumbers between $2N_\eps-3\delta/\eps$ and $2N_\eps-3\delta/\eps$.
By choosing $\delta < 2/3$, say $\delta = 1/3$, we thus guarantee the existence of constants
$C$ and $c$ independent of $\eps$ such that
\begin{equ}
\bigl\|e^{t\Delta_{\eps}} (\Pi_{\delta/\eps}A)^3 e_{2N_\eps}\bigr\|_{\cC^0_a}
\le C \eps^{-1} e^{-c\eps^{-2}t} \bigl\|(\Pi_{\delta/\eps}A)^3\bigr\|_{\cC^0_a}\;.
\end{equ}
Hence,
\begin{equs}
 \Bigl\| \int_0^t \pi_\eps e^{(t-\tau)\Delta_{\eps}} &
\left( \left( \Pi_{\delta/\eps} A(\tau) \right)^3 e^{\frac{2 i \pi N_\eps x}{L}}
\right) d \tau \Bigr\|_{\cC^0_u}
  \\ & \leq
     C \int_0^t e^{-c\eps^{-2}(t-\tau)}\eps^{-1}
    \|  \Pi_{\delta/\eps} A(\tau)  \|_{\cC^0_a}^3 d \tau
       \\ & \leq
     C \eps \sup_{t \in[0,T_0]} \|\Pi_{\delta/\eps}A(t) \|_{\cC_a^0}^p\;.
 \label{e:thm34b}
\end{equs}
Since furthermore $\|\pi_\eps e^{t\Delta_{\eps}} \|_{\cL(\cC_a^0, \cC_u^0)}
\leq C$ independently of $\eps$, we obtain:
\begin{equ}[e:thm34a]
 \Bigl\| \int_0^t \pi_\eps e^{(t-\tau)\Delta_{\eps}}
\Bigl( \bigl( \Pi^c_{\delta/\eps} A(\tau) \bigr)^3 e^{\frac{2 i \pi N_\eps x}{L}}
\Bigr) d \tau \Bigr\|_{\cC^0_u}
 \leq
     C \sup_{t \in[0,T_0]} \|\Pi^c_{\delta/\eps}A(t) \|_{\cC_a^0}^p\;.
\end{equ}
Combining \eref{e:thm34b}, \eref{e:thm34a}, and \eref{e:lastone}, we obtain
\begin{equ}
\E\sup_{t \in [0,T]} \|  I_1(t)\|_{\cC_u^0}^p \leq C_p \eps^{\frac{p}{2}}.
\end{equ}
Putting all the above estimates together we obtain \eqref{e:est_resid}
of Theorem \ref{thm:res}.
\end{proof}
%
%
%
\section{Main Approximation Result}
\label{sec:approximation}
This section is devoted to the proof of the following approximation theorem.
\begin{theorem}{\bf (Approximation)}
\label{thm:approx}
Fix $T_0>0$, $p \ge 1$, and $\kappa >0$.
There exist joint realisations of the Wiener
 processes $W_\xi$ and $W_\eta$ from \eref{e:stochconv}
such that, for every admissible initial condition $A(0)$,
 there exists $C > 0$ such that
\begin{equ}
\mathbb{E} \Big( \sup_{t \in [0, T_0]} \| u(t) - \pi_\eps A(t)\|^{p}_{\cC^0_u}\Big) \leq C
\eps^{\frac{p}{2} - \kappa} \;.
\label{e:c0_est}
\end{equ}
where $A$ is the solution of \eref{e:Avoc} with initial condition $A(0)$
and $u$ is the solution of \eref{e:uvoc}
with initial condition $u(0)= \pi_\eps A(0)$.
\end{theorem}

Before we turn to the proof of this result, we make a few preliminary calculations.
Let $A(t)$ and $u(t)$ be as in the statement of \theo{thm:approx} and define
$$R(t)= u(t)- \pi_\eps A(t)\;.
$$
From \eref{e:uvoc} and Definition~\ref{def:residual} we easily derive
\begin{equs}
R(t) &=  \int_0^t e^{(t-\tau)\cL_{\epsilon}}[\tilde\nu R(\tau)-3 R(\tau) (\pi_\eps A(\tau))^2
-3 R(\tau)^2\pi_\eps A(\tau)- R(\tau)^3]d\tau \\
 &\ \qquad\qquad + \mathrm{Res}(\pi_\eps A)(t).
\nonumber
\end{equs}
Define $$\phi(t) = \mathrm{Res}(\psi)(t), \quad \psi(t) = \pi_\eps A(t)$$ and
\begin{equ}[e:rdef]
r(t) = R(t) - \phi(t).
\end{equ}
Then $r(t)$ satisfies the equation
\begin{equation}
\partial_t r = \cL_{\epsilon}r +  \tilde \nu (r + \phi) - 3 (r + \phi) \psi^2 - 3
(r + \phi)^2 \psi -   (r + \phi) ^3, \quad r(0) = 0.
\label{e:r}
\end{equation}
With these notations, we have the following \textit{a priori} estimates in $L^2$.
\begin{lemma}
\label{lem:l2_est}
Under the assumptions of Theorem \ref{thm:approx}
there exists a constant $C>0$
such that
\begin{equ}
\mathbb{E} \Big( \sup_{t \in [0, T_0]} \| r(t)\|^{p}\Big) \leq C \eps^{\frac{p}{2}
- \kappa}\;,
\label{e:l2_est}
\end{equ}
for $r(t)$ defined in \eqref{e:rdef}.
\end{lemma}

\begin{proof}
As before, we are using $\| \cdot \|$ to denote the norm in  $\cH_u$. We take the scalar product of (\ref{e:r}) with $r$ to obtain
\begin{equs}
\frac{1}{2} \frac{d}{d t} \|r \|^2
& =  \langle \cL_{\epsilon} r, r \rangle
+ \tilde\nu \langle r
+ \phi , r \rangle
- 3  \langle (r + \phi) \psi^2 , r \rangle
- 3  \langle (r + \phi)^2 \psi , r \rangle
  \\ &
\quad - \langle (r + \phi)^3  , r \rangle \\ & =:
           I_1 + I_2 + I_3 + I_4 + I_5\;.
\end{equs}
Since $\cL_{\epsilon}+1$ is by definition a
non-positive selfadjoint operator, we have $I_1 \leq  - \| r \|^2$.
Moreover, the Cauchy-Schwarz inequality yields:
$$
I_2 \leq C \| r \|^2 + C \| \phi\|^2\;.
$$
It follows from the Young and Cauchy-Schwarz inequalities that
$$
I_3 \leq - 3 \int_{-L}^L r^2 \psi^2 \, dx + C \| r\|^2 + C \|\phi \|^2_{\cC_u^0} \|
\psi\|_{\cC_u^0}^4\;,
$$
and
\begin{equs}
I_4 & =  - 3  \int_{-L}^L r^3 \psi \, dx - 6  \int_{-L}^L
r^2 \phi \psi \, dx - 3  \int_{-L}^L r \phi^2 \psi \, dx
      \\ & \leq
     \frac18 \|r \|^4_{L^4_u} + C \|\psi \|^4_{\cC_u^0}+C \|\phi \|^2_{\cC_u^0} \|\psi \|^2\;.
\end{equs}
Finally, expanding $I_5$ yields
$$
I_5 \leq - \frac78 \|r \|^4_{L^4_u} + C  \| \phi\|^4_{\cC_u^0}\;.
$$
Putting all these bounds together,  we obtain:
$$
\partial_t\| r \|^2 \leq C \| r \|^2 + C \left(1 + \|\psi \|^4_{\cC_u^0} \right)
\|\phi \|^2_{\cC_u^0} \left(1 + \|\phi \|^2_{\cC_u^0} \right)\;.
$$
We apply now a comparison argument to deduce ($r(0)=0$ by definition)
\begin{equation}
\| r(t) \|^2 \leq  C\int_0^t e^{C(t-\tau)}
 \left(1 + \|\psi \|^4_{\cC_u^0} \right)\|\phi \|^2_{\cC_u^0}
\left(1 + \|\phi \|^2_{\cC_u^0} \right) (\tau)d\tau.
\label{e:r_ineq}
\end{equation}
From Theorem \ref{thm:res} we derive with $\phi(t) = \mbox{Res}(\pi_\eps A)(t)$
\begin{equation}
\label{e:apriphi}
\mathbb{E}  \sup_{t \in [0, T_0]} \|  \phi (t)\|^{p}_{\cC^0_u} \leq C_p
\eps^{\frac{p}{2} - \kappa}\;.
\end{equation}
Furthermore, the \textit{a priori} estimate on $A(t)$, Proposition \ref{prop:apriori}, together with
the properties of $\pi_\eps$ yield for $\psi(t) = \pi_\eps A(t)$
\begin{equation}
\label{e:apripsi}
\mathbb{E}  \sup_{t \in [0, T_0]} \|  \psi (t)\|^{p}_{\cC^0_u} \leq C_p\;.
\end{equation}
Combining \eqref{e:r_ineq} with \eref{e:apriphi} and \eref{e:apripsi}
we obtain \eqref{e:l2_est} of Lemma \ref{lem:l2_est}.
\end{proof}
\bigskip
To proceed further we first establish two interpolation inequalities.
We start by defining the selfadjoint operator
\begin{equation}
\cA = \iota_\eps^* (1 - \partial_x^2) \iota_\eps\;.
\label{e:opa}
\end{equation}
By Definition \ref{def:sobolev}, the $\CH_u^\alpha$-norm
is given by
 $\|r \|_\alpha = \langle r, \cA^\alpha r \rangle$.
Furthermore, the following
interpolation lemma holds.
\begin{lemma}
\label{lem:interp}
For $p \geq 2$ there is a constant $C>0$ such that
$$\|u\|_{L^p_u}
    \le C  \|u\|_1^{\frac12-\frac1p} \|u\|^{\frac12+\frac1p}
    \quad\mathrm{and}\quad
    \|u\|_{L^p_u}
    \le C  \| u\|_2^{\frac14-\frac1{2p}} \|u\|^{\frac34+\frac1{2p}}$$
for every $u \in \cH^2_u$.
\end{lemma}
\begin{proof}
The proof of the lemma follows from the standard interpolation inequalities,
the definition of $\cA$ and the properties
of the operators $\iota_\eps, \, \pi_\eps$ (cf. \eref{e:defpi} and \eref{e:defiota}).
\end{proof}
It is also straightforward to verify
that $\cL_\eps$ and $\cA$ have a joint basis of eigenfunctions
consisting of $\sin(\pi kx/L)$ and $\cos(\pi kx/L)$.
By comparing the eigenvalues it is easy to verify that
\begin{equation}
\langle -\cL_\eps u, u \rangle \geq \langle \cA u, u \rangle
 \quad\mathrm{and\ thus}\quad \|u \|_1 \leq
\|(-\cL_\eps)^{\frac{1}{2}} u \|\;.
\label{e:l_est1}
\end{equation}
Furthermore
\begin{equation}
\langle -\cL_\eps u, \cA u \rangle \geq  \| \cA u \|^2 = \|u \|_2^2\;.
\label{e:l_est2}
\end{equation}

We now turn to the
\begin{proof}[of Theorem \ref{thm:approx}]
 We take the scalar product of (\ref{e:r}) with $\cA r$ to obtain
\begin{equs}
\frac{1}{2} \partial_t \| r \|_1^2 & =  \langle \cL_{\epsilon} r, \cA r \rangle +
\tilde\nu \langle r + \phi , \cA r \rangle - 3  \langle (r + \phi) \psi^2 , \cA r \rangle
 \\ & \quad
- 3  \langle (r + \phi)^2 \psi , \cA r \rangle
- \langle (r + \phi)^3  , \cA r \rangle
       \\ & =:
           I_1 + I_2 + I_3 + I_4 + I_5\;.
\end{equs}
We then use \eqref{e:l_est2} to get $I_1 \leq  - \| r \|_2^2$.
Moreover, using Cauchy-Schwarz and Young, one has the bounds
$$
I_2 \leq C \| r \|^2 + C \| \phi\|^2 + \frac18 \| r \|_2^2
$$
and
$$
I_3 \leq  C \| r\|^2 \|\psi \|^4_{\cC^0_u}  + C
\|\phi \|^2 \| \psi\|_{\cC^0_u}^4 + \frac18 \|r \|_2^2\;.
$$
In order to bound the term $I_4$ we use Lemma \ref{lem:interp} with $p = 4$:
\begin{equ}
I_4 =  \frac18\| r \|_2^2 +
     C \| \psi \|_{\cC^0_u}^{\frac{8}{3}}
     \|r \|^{\frac{14}{3}}
    + C  \|\psi \|^2_{\cC^0_u} \| \phi\|^4_{\cC^0_u}\;.
\end{equ}
Finally, we use Lemma \ref{lem:interp} with $p = 6$ to bound $I_5$:
$$
I_5 \leq \delta \| r \|_2^2 + C_{\delta} \|\phi \|^6_{\cC^0_u}
+ C_{\delta} \| r\|^{10}\;.
$$
Putting everything together  we obtain:
\begin{equa}
\partial_t \| r \|_1^2 & \leq C \| r \|^2 \left( \|\psi \|^4_{\cC^0_u} +
 \|\psi \|^3_{\cC^0_u} \|r \|^2 + \|\psi \|^2_{\cC^0_u} \|r \|^4+  \|r \|^8 \right)
  \\ &\quad
 +
C \|\phi \|^2_{\cC^0_u} \left(1 + \|\phi \|^2_{\cC^0_u}
\|\psi \|^2_{\cC^0_u} + \|\psi \|^4_{\cC^0_u} + \|\phi \|^4_{\cC^0_u}
\right)\;.
\label{e:r*}
\end{equa}
Estimate \eqref{e:c0_est} follows now from \eqref{e:r*},  together with Lemma \ref{lem:l2_est} and the \textit{a priori}
bounds on $\phi$ and $\psi$ from \eref{e:apripsi} and \eref{e:apriphi}.
\end{proof}
%
%
%
\section{Attractivity}
\label{sec:attr}
This section provides attractivity results for the SPDE.
We consider the rescaled equation \eref{e:orig2},
and we prove that regardless of the initial condition $u(0)$ we start with,
we will end up  for sufficiently large $t>0$  with an admissible $u(t)$, thus giving
admissible initial conditions for the amplitude equation. The main result of
this section is contained in the following theorem.
\begin{theorem}{\bf (Attractivity)}
\label{thm:admu}
For all (random) initial conditions $u(0)$ such that $u(0) \in \cH_u$ almost surely
 and every $t >0$, the mild
solution $u(t)$
of \eref{e:orig2} is admissible in the sense of Definition \ref{def:admA}. Furthermore, given a
$T_0 >0$ the family of constants $\{ C_q \}_{q>0}$  which appears in the definition of
admissibility is independent of the initial conditions and the time $t$ for $t > T_0$.
\end{theorem}

\begin{remark}
In \cite{MR2000b:60157} and \cite{MR2002h:60118}
uniform bounds on the solutions after transient times
were obtained that are independent of the initial condition.
 However, the statements given in these papers
do not cover the situation presented here.
\end{remark}

The rest of this section is devoted to the proof of this theorem.
First we will prove standard a-priori estimates in
$L^2$-spaces that rely on the strong nonlinear stability of the equation. Then we will provide
regularisation results using the $\cH^1_u$ norm which allow us to get to the $\cC_u^0$ space
and we end with the admissibility of the solution.
Note that the solution $u$ will never be in $\cH^1$,
therefore we have to consider suitable transformations.

Let $u(t)$ denote the mild solution of \eqref{e:orig2}, \ie a solution of
\eref{e:uvoc}.
Denote as in \eref{e:stochconv1}
by $ W_{\CL_{\eps}}$ the stochastic convolution for the operator $\CL_{\eps}$
and define $v := u - W_{\CL_{\eps}}$. Then $v$ satisfies the equation
\begin{equation}
\partial_t v = \CL_{\epsilon} v +  \tilde\nu (v + W_{\cL_{\eps}}) - (v +  W_{\cL_{\eps}})^3,
\label{e:v}
\end{equation}
with the same initial conditions as $u$. We start by obtaining an $L^2$ estimate on
$u$. Before we do this let us discuss some estimates for the stochastic convolution.
Using first Proposition \ref{theo:stoch_conv}
we obtain
$$ \EX  \sup_{t\in[0,T_0]}\| W_{\cL_\eps}(t)\|^{2p}_{\cC^0_u}
\le C\EX  \sup_{t\in[0,T_0]}\|{W}_{\Delta_\eps}(t)\|^{2p}_{\cC^0_a}
+ C \eps^{p/2 - \kappa}\;.
$$
Hence, using the modification
 of Lemma \ref{lem:APBphi} or Proposition \ref{prop:apriori} with $c=0$,
\begin{equ}[e:attscest]
\EX  \sup_{t\in[0,T_0]}\| W_{\cL_\eps}(t)\|^{2p}_{\cC^0_u} \le C\;.
\end{equ}
\begin{lemma}
\label{lem:u_l2}
Let $u(t)$ be the solution of \eqref{e:uvoc}. Fix arbitrary $T_0>0$. Then  there
exists a constant $C>0$ independent of $u(0)$ such that
$$ \sup_{t\ge T_0} \EX \|u(t) \|^p \le C.$$
Assume further that  $\EX \|u(0)\|^p \le c_0$. Then, given $T_0>0$ there
exists a constant $C$ such that
$$
 \sup_{t\ge0} \EX \|u(t) \|^p \le C, \quad\mathrm{and}\quad
 \EX  \sup_{t\in[0,T_0]} \|u(t) \|^p \le C.
$$
\end{lemma}
\proof We multiply \eqref{e:v} with $v$, integrate over $[-L,L]$, use the dissipativity of
$\CL_{\eps}$ in $\cH_u$, together with the fact that
$$-\langle v, (v +  W_{\cL_{\eps}})^3 \rangle
\leq - (1 - \delta) \| v\|^4 + \delta \| v \|^2 + C_{\delta} \| W_{\cL_{\eps}} \|^4$$
for every $\delta >0$, which we choose to be sufficiently small, to obtain
$$
\partial_t \|v \|^2  \leq - C_1 \|v \|^4
    + C_2 \left(1 + \| W_{\CL_{\eps}} \|^4_{\cC_u^0}\right),
$$
for some positive constants $C_1$ and $C_2$. A comparison theorem for ODE yields
for $t \in[0,T_0]$
\begin{equs}
\|v(t) \|^2
&\leq \max\Big\{ C(1+ \sup_{t\in[0,T_0]}\| W_{\CL_{\eps}} \|^2_{\cC_u^0})
                ; \frac{1}{C_1 t /2 + 1/\|v(0)\|^2}\Big\}\\
&\leq C \Big(1+\sup_{t\in[0,T_2]}\| W_{\CL_{\eps}} \|^2_{\cC_u^0}+\frac1t  \Big)\;.
\label{e:comp}
\end{equs}
Note furthermore, that
$$
\partial_t \|v \|^2     \leq - c \|v \|^2 + C \left(1 + \| W_{\CL_{\eps}} \|^4_{\cC^0_u}\right)\;.
$$
Again a comparison argument for ODEs yields
for any $T_0>0$
\begin{equ}[e:comp2]
\|v(t) \|^2 \le
e^{c(t-T_0)}\|v(T_0)\|^2
+ C\int_{T_0}^t e^{-c(t-s)} \left(1 + \| W_{\CL_\eps}(s) \|^4_{\cC_u^0}\right)ds
\end{equ}
The claims of the lemma follow now easily from \eqref{e:comp} and \eqref{e:comp2}, the
fact that $u = v + W_{\CL_\eps}$, and the estimates on the stochastic convolution
from \eref{e:attscest}. \qed
\begin{lemma}
\label{lem:u_apriori}
Fix $\delta>0$, $p>0$, and $T_0>0$. Then there is a constant $C$ such that
for all  mild the solutions $u$ of \eqref{e:orig2} (\ie\ \eqref{e:uvoc}) with
$\E \|u(0) \|^{5p} \leq \delta$
the following estimate holds
\begin{equation}
 \sup_{t \geq T_0} \EX \|u(t) \|_{\cC^0_u}^p \leq C\;.
\label{e:c0_u}
\end{equation}
%
\end{lemma}
\proof
Define
$$
w(t):=u(t)-e^{t\cL_\eps}u(0)- W_{\CL_{\eps}} =: u(t)-\varphi(t)
$$
Now $w$ fulfills
\begin{equation}
\partial_t w = \CL_{\epsilon} w +  \tilde\nu (w + \varphi) - (w +\varphi)^3,
\quad w(0)=0
\label{e:defw}
\end{equation}
Consider $\cA$ defined in \eref{e:opa} and
multiply \eref{e:defw} with $\cA w$, integrate over $[-L, L]$, use Lemma
\ref{lem:interp} with $p=6$ as well as $\|v \|_1 \leq \| v \|_2 $
to obtain:
$$
\partial_t \|w \|_1^2 \leq - C_1 \|w\|_1^2 + C_2 \left( \|w \|^2 + \|w \|^{10} +
\|\varphi \|^2 + \| \varphi \|_{L_u^6}^6 \right)
$$
A comparison theorem for ODE now yields:
\begin{equ}
\|w(t) \|_1^2 \leq  C_2 \int_0^t e^{-C_1 (t - \tau)} (1+\|w \|^{10}+\|\varphi \|_{L_u^6}^6)(\tau) \, d \tau\;.
\label{e:v_h*}
\end{equ}
Using \eref{e:l_est1} and Lemma \ref{lem:interp} we deduce that
$\|u\|_{L^6_u} \le C \|(-\cL_\eps)^{1/2} u\|^{1/3}\|u\|^{2/3}$.
Hence,
\begin{equ}[e:SGb3]
\|e^{t\cL_\eps}u_0\|_{L^6_u}^3 \le C t^{-1/2}\|u_0\|^3\;.
\end{equ}
Taking the $\cL^{p/2}$-norm in probability space, we deduce from \eqref{e:v_h*}
using \eref{e:SGb3} and the embedding of $\cH^1_u$ into $\cC_u^0$
from Remark \ref{rem:H1intoC0}
\begin{equs}
\Big(\EX \|w(t) \|_{\cC^0_u}^p\Big)^{2/p}
& \leq  C \left(1
+ \sup_{t \geq 0} ( \E \|w(t) \|_{\cC^0_u}^{5p} )^{2/p}
+ \sup_{t \geq 0} \E ( \| W_{\cL_\eps}\|^{3p}_{L_u^6}\Big)^{2/p} \right)
 \\&\quad
+ C\int_0^t \tau^{-1/2} e^{-C_1\tau}d\tau (\EX \|u(0)\|^{3p})^{2/p}
  \leq  C
\label{e:est_w}
\end{equs}
for all $t>0$, where we used the $L^2$-bounds from Lemma \ref{lem:u_l2}.
Note that this is the reason, why we need the $5p$-th moment of the
initial condition $u(0)$.
On the other hand, the bound on the stochastic convolution together with standard properties of
 analytic semigroups enable us to bound $\varphi(t)$, for $t$ sufficiently large:
$$
\|\varphi(t) \|_{\cC^0_u} \leq C \|e^{t \cL_\eps} u(0) \|_1 + \| W_{\cL_\eps} \|_{\cC^0_u}
\le Ct^{-1/2}\|u(0)\|+\| W_{\cL_\eps} \|_{\cC^0_u}.
$$
Estimate \eqref{e:c0_u} now follows from the above estimate,
Lemma \ref{lem:u_l2}, the definition of $w$ and estimate \eqref{e:est_w}.
\qed
\proof[of Theorem \ref{thm:admu}]
First, Lemma \ref{lem:u_l2} together with Lemma \ref{lem:u_apriori} establishes
the existence of a time $T_0>0$
such that $\EX\|u(t)\|^p_{\cC_u^0}\le C$ for all $t\ge T_0$.
Furthermore, combining \eqref{e:v_h*} and \eref{e:est_w}
we immediately get that
$$
\E \|w(t) \|_{1}^p \leq C.
$$
Thus, under the assumptions of the previous lemma and using the properties of the
stochastic convolution $ W_{\CL_\eps}(t)$ we conclude that for every $t > 0$
$u(t)$ can be decomposed as
$$
u(t) = w(t) + Z(t)+e^{t\cL_\eps}u(0)\;,
$$
where $w(t) \in \CH^1_u$ and $Z(t)$ is a centred Gaussian process in
$\cC^0_u$. Moreover, $e^{t\cL_\eps}u(0)$ is in $\CH_u^1$
for any $t>0$, too. We use now the decomposition
$$
u(T_0+\tau) = \tilde w(\tau) + \tilde Z(\tau)+e^{\tau\cL_\eps}u(T_0)\;,
$$
where we consider $u(t)$
as the solution starting at
sufficiently large $T_0>0$ with initial conditions $u(T_0)$.
For $\tau>0$ sufficiently
large the process
$\iota_\eps \tilde Z(\tau) := \iota_\eps W_{\cL_\eps}(\tau)$ (in law)
is clearly as in $2.$ of Definition
\ref{def:admA}.
For 1. define $W_0(\tau):=\tilde w(\tau) + e^{\tau\cL_\eps}u(T_0)$.
We obtain from Lemma \ref{lem:u_apriori}
and the analog of \eref{e:est_w} for $\tilde w$  that
$$\EX\|W_0(\tau)\|_1^p \le C_p + C \tau^{-p/2}\EX \|u(T_0) \|^p \le C\;.
$$
Hence, the decomposition $u(t)=W_0(t-T_0)+\tilde Z(t-T_0)$
shows the admissibility of $u(t)$,
where the constants are independent of $t\ge 2T_0$. \qed

%
%
\section{Approximation of the Invariant Measure}
\label{sec:inv_meas}
First, we denote by $\CP_t^\eps$ the semigroup (acting on finite Borel measures) associated
to \eref{e:orig2} and by $\CQ_t^\eps$ the semigroup associated to \eref{e:ampl2}.
Note that $\CQ_t^\eps$ depends on $\eps$,
but it is for instance independent of $\eps$ for $L\in \eps\pi\N$.

Recall also that
the Wasserstein
distance $\|\cdot\|_W$ between two measures on some metric space $\CM$
with metric $d$ is given by
\begin{equ}
\|\mu_1-\mu_2\|_W = \inf_{\mu \in \CC(\mu_1,\mu_2)} \int_{\CM^2} \min\{1, d(f,g)\}\,\mu(df,dg)\;.
\end{equ}
where $\CC(\mu_1,\mu_2)$ denotes the set of all measures on $\CM^2$ with
$j$-th marginal  $\mu_j$.
See for example \cite{rachev91} for detailed properties of this distance.

In the sequel, we will use the
notation $\|\mu_1-\mu_2\|_{W,p}$ for the Wasserstein distance corresponding to the $\L^p$-norm
$d(f,g) = \|f-g\|_p$
for $p \in [1,\infty]$. The main result on the invariant measures is
\begin{theorem}\label{theo:IM}
Let $\mu_{\star,\eps}$ be an invariant measure for \eref{e:orig2} and let $\nu_{\star,\eps}$
 be the
(uni\-que) invariant measure for \eref{e:ampl2}. Then, for every $\kappa > 0$, there exists $C>0$
such that one has
\begin{equ}
\|\mu_{\star,\eps} - \pi_\eps^* \nu_{\star,\eps}\|_{W,\infty} \le C \eps^{1/2-\kappa}
\end{equ}
for every $\eps \in (0,1]$.
\end{theorem}

Note that $\nu_{\star,\eps}$  is actually independent of $\eps$ provided $L\in \eps\pi\N$.
As usual, the measure $\pi_\eps^* \nu$ denotes the distribution of
$\pi_\eps$ under the measure $\nu$.

\begin{proof}
Fix $\kappa > 0$ for the whole proof.
From the triangle inequality
and the definition of an invariant measure,
we obtain
\begin{equa}[e:startrel]
\|\mu_{\star,\eps} - \pi_\eps^* \nu_{\star,\eps}\|_{W,\infty}
&\le \|\CP_t^\eps \mu_{\star,\eps} - \pi_\eps^* \CQ_t^\eps \iota_\eps^* \mu_{\star,\eps}\|_{W,\infty}\\
&\quad + \|\pi_\eps^* \CQ_t^\eps \nu_{\star,\eps} - \pi_\eps^* \CQ_t^\eps \iota_\eps^*\mu_{\star,\eps}\|_{W,\infty} \;.
\end{equa}

Concerning the first term, it follows from Theorem \ref{thm:approx}
that the family of measures $\mu_{\star,\eps}$ is admissible and
that
\begin{equ}
\|\CP_t^\eps \mu_{\star,\eps} - \pi_\eps^* \CQ_t^\eps \iota_\eps^* \mu_{\star,\eps}\|_{W,\infty}
\le C \eps^{1/2 - \kappa}\;.
\end{equ}

In order to bound the second term in \eref{e:startrel}, we use the
exponential convergence of $\CQ_t^\eps \mu$ towards a unique invariant measure.
This is a well-known result for SPDEs driven by space-time white noise
(cf. \eg Theorem 2.4 of \cite{MR2002h:60118}),
but we need the explicit dependence of the constants on the
initial measures. The precise bound required for our proof is given in \lem{lem:almostgood} below.

By \lem{lem:almostgood}, there exists $t > 0$ such that
\begin{equ}
\|\CQ_t^\eps \mu_{\star,0} - \CQ_t^\eps \iota_\eps^*\mu_{\star,\eps}\|_{W,\infty}
\le {1\over 2 \sqrt L} \|\iota_\eps^* \mu_{\star,\eps} - \nu_{\star,\eps}\|_{W,2}\;,
\end{equ}
so that the boundedness in $\L^\infty$ of $\pi_\eps$ implies
\begin{equ}
\|\mu_{\star,\eps} - \pi_\eps^* \nu_{\star,\eps}\|_{W,\infty}
\le {1\over 2 \sqrt L} \|\iota_\eps^* \mu_{\star,\eps} - \nu_{\star,\eps}\|_{W,2} + C \eps^{1/2-\kappa}\;.
\end{equ}
Since the $\L^2$-norm is bounded by $\sqrt L$ times the $\L^\infty$-norm, this in turn is smaller than
\begin{equ}
{1\over 2} \|\mu_{\star,\eps} - \pi_\eps^* \nu_{\star,\eps}\|_{W,\infty}
+ {1\over 2 \sqrt L} \|\iota_\eps^*\pi_\eps^* \nu_{\star,\eps} - \nu_{\star,\eps}\|_{W,2}
+ C \eps^{1/2-\kappa}\;.
\end{equ}
It follows from standard energy-type estimates that
\begin{equ}
\expect \int_{\cH_a^\alpha} \|A\|_\alpha  \,\nu_{\star,\eps}(dA) < C_\alpha
\end{equ}
for every $\alpha < 1/2$, where the constants $C_\alpha$ can be chosen independently
of $\eps$. This estimate is a straightforward extension of the results
presented in Section \ref{sec:apeftae}.

One therefore has
$ \|\iota_\eps^*\pi_\eps^* \nu_{\star,\eps} - \nu_{\star,\eps}\|_{W,2}
\le C_\kappa \eps^{1/2-\kappa}$.
Plugging these bounds back into \eref{e:startrel} shows that
\begin{equ}
\|\mu_{\star,\eps} - \pi_\eps^* \nu_{\star,\eps}\|_{W,\infty}
\le {1\over 2}\|\mu_{\star,\eps} - \pi_\eps^* \nu_{\star,\eps}\|_{W,\infty}
+ C_\kappa \eps^{1/2-\kappa}\;,
\end{equ}
and therefore concludes the proof of \theo{theo:IM}.
\end{proof}
Besides the approximation result, the main ingredient for the above reasoning is:

\begin{lemma}\label{lem:almostgood}
For every  $\delta > 0$, there exists a time $T=T(\delta)$ independent of $\eps$ such that
\begin{equ}
\|\CQ_T^\eps \mu - \CQ_T^\eps \nu\|_{W,\infty} \le \delta \|\mu - \nu\|_{W,2}\;.
\end{equ}
\end{lemma}
\begin{proof}
It follows from the Bismut-Elworthy-Li formula combined with standard \textit{a priori} bounds
on $\CQ_t^\eps$ \cite{XueMei,ZDP,MR2000b:60157} that
\begin{equ}
\|\CQ_t^\eps \mu - \CQ_t^\eps \nu\|_{TV} \le C \bigl(1 + t^{-1/2}\bigr) \|\mu - \nu\|_{W,2}\;,
\end{equ}
with a constant $C$ independent of $\eps$.

On the other hand, \cite{MR2002h:60118} there exist constants $C$ and $\gamma$ such that
\begin{equ}[e:boundexp]
\|\CQ_t^\eps \mu - \CQ_t^\eps \nu\|_{TV} \le C e^{-\gamma t} \|\mu - \nu\|_{TV}\;.
\end{equ}
These constants may in principle depend on $\eps$. By retracing the constructive
argument of Theorem~5.5 in  \cite{MR1939651} with the binding function
\begin{equ}
G(x,y) = -C(y-x)\bigl(1 + \|y-x\|^{-1/2}\bigr)\;,
\end{equ}
one can however easily show that the constants in \eref{e:boundexp} can be chosen independently
of $\eps$.
\end{proof}
%
%
%
\section{Approximation of the Stochastic Convolution}
\label{sec:stoch_conv}
In this section, we give $\L^\infty$ bounds in time and in space on the difference between the
stochastic convolutions of the original equation and of the amplitude equation.
The main result of this section is
\begin{theorem}\label{theo:stoch_conv}
Let $W_{\CL_\eps}$ and $W_{\Delta_\eps}$ be defined as in \eref{e:stochconv}, and let the correlation functions $q_\eps$
 with Fourier coefficients $q^\eps_k$ satisfy Assumptions~\ref{ass:g} and \ref{ass:gkeps} below.
For every $T>0$, $\kappa > 0$, and $p \ge 1$ there exists a constant $C$ and a joint realisation of $W_{\CL_\eps}$ and $W_{\Delta_\eps}$ such that
\begin{equ}
\expect \Big( \sup_{t \in [0,T]} \|W_{\CL_\eps}(t) - \pi_\eps W_{\Delta_\eps}(t)\|_{\CC^0_u}^p \Big)
\le C \eps^{{p\over 2}-\kappa}\;,
\end{equ}
for every $\eps \in (0,1)$.
\end{theorem}
We will actually prove a more general result, see \prop{prop:stoch_conv} below, which has
\theo{theo:stoch_conv} as an immediate corollary. The general result allows the linear operator $\CL_\eps$ to be essentially an arbitrary real differential operator instead of restricting it to the operator $-1-\eps^{-2}(1+\eps^2\d_x^2)^2$.
Our main technical tool is a series expansion of the stochastic convolution
together with \lem{lem:Dirk},
which will be proved in Section~\ref{subsec:ser_exp} below.
The expansion with respect to space is performed using Fourier series.
For the expansion in time we do not use Karhunen-Loeve expansion directly, since we
do not necessarily need an orthonormal basis to apply \lem{lem:Dirk}. Our choice of an appropriate basis
will simplify the coefficients
in the series expansion significantly (cf. Lemma \ref{lem:KLexp}).
We start by introducing the assumptions required for the differential operator $P(i\partial_x)$.
\begin{assumption} \label{ass:P}
Let $P$ denote an even function $P: \R \to \R$ satisfying the
following properties:
\begin{claim}
\item[\textbf{P1}] $P$ is three times continuously differentiable.
\item[\textbf{P2}] $P(\zeta) \ge 0$ for all $\zeta \in \R$ and $P(0) > 0$.
\item[\textbf{P3}] The set $\{\zeta\,|\, P(\zeta) = 0\}$ is finite and will be denoted by $\{\pm\zeta_1,\ldots,\pm\zeta_m\}$. Note that $\xi_j\not=0$.
\item[\textbf{P4}] $P''(\zeta_j) > 0$ for $j=1,\ldots,m$.
\item[\textbf{P5}] There exists $R>0$ such that $P(\zeta) \ge |\zeta|^2$ for all $\zeta$ with
$|\zeta| \ge R$.
\end{claim}
\end{assumption}
Note that choosing $P$ even ensures that $P(i\partial_x)$ is a real operator,
but our results also hold for non-even $P$, up to trivial notational complications.

We now make precise the assumptions on the noise that drives our equation.
Consider an even real-valued distribution $q$ such that its Fourier transform satisfies $\hat q \ge 0$.
Then, $q(x)\delta(t)$ is the correlation function for a real distribution-valued Gaussian process
$\xi(x,t)$ with $x,t \in \R^2$, \ie a process such that $\expect \xi(s,x) \xi(t,y) = \delta(t-s) q(x-y)$.
We restrict ourself to correlation functions in the following class:
\begin{assumption} \label{ass:g}
The distribution $q$ is such that
$\hat{q}\in L^\infty(\R)$ and $\hat{q}$
is globally Lipschitz continuous.
\end{assumption}
At this point, a small technical difficulty arises from the fact that we want to replace
$\xi$ by a $2L/\eps$-periodic translation invariant noise process $\xi_\eps$ which
is close to $\xi$ in the bulk of this interval. Denote by $q^\eps$ the $2L/\eps$-periodic
correlation function of $\xi_\eps$ and by $q_k^\eps$ its Fourier coefficients, \ie
\begin{equ}[e:Dirk]
q_k^\eps = \int_{-L/\eps}^{L/\eps} q^\eps(x)\, e^{-i{k\pi\eps\over L} x}\,dx\;.
\end{equ}
One natural choice is to take for $q^\eps$ the periodic continuation of the restriction
of $q$ to $[-L/\eps, L/\eps]$. This does however not guarantee that $q^\eps$ is again
positive definite. Another natural choice is to define $q^\eps$ via its Fourier coefficients by
\begin{equ}[e:Martin]
q_k^\eps = \int_{-\infty}^{\infty} q(x) \, e^{-i{k\pi\eps\over L} x}\,dx\;,
\end{equ}
which corresponds to taking $q^\eps(x) = \sum_{n \in \Z} q(x + 2nL/\eps)$. This guarantees
that $q^\eps$ is automatically positive definite, but it requires some summability of $q$.
Note that for noise with bounded correlation length (\ie support of $q$ uniformly bounded)
\eref{e:Dirk} and \eref{e:Martin} coincide for $\eps>0$ sufficiently small.

We choose not to restrict ourselves to one or the other choice, but to impose only
a rate of convergence of the coefficients $q_k^\eps$ towards $\hat{q}(k\pi\eps/L)$:
\begin{assumption} \label{ass:gkeps}
Let $q$ be as in Assumption \ref{ass:g}.
Suppose there is a non-negative approximating sequence $q_k^\eps$
that satisfies
\begin{equ}
\sup_{k\in\N_0}|\sqrt{q_k^\eps}-\sqrt{\hat{q}(k\pi\eps/L)}|\le C\eps\;,
\end{equ}
for all sufficiently small $\eps>0$.
\end{assumption}
\begin{example}
A simple example of noise fulfilling Assumptions \ref{ass:g} and \ref{ass:gkeps}
is given by space-time white noise.  Here $\hat q(k)=1$ and the natural approximating sequence
is $q_k^\eps=1$ for all $k$.
\end{example}

A more general class of examples is given by the following lemma.
\begin{lemma} \label{lem:hatgexp}
Let $q$ be positive definite and such that $x \mapsto (1 + |x|^2)\,q(x)$ is in $\L^1$.
Define $q_k^\eps$ either
by \eref{e:Martin} or by \eref{e:Dirk} (in the latter case, we assume additionally that the
resulting $q^\eps$ are positive definite). Then Assumptions~\ref{ass:g} and \ref{ass:gkeps}
are satisfied.
\end{lemma}
\begin{proof}
This follows from elementary properties of Fourier transforms.
\end{proof}

Let us now turn to the stochastic convolution, which is the solution to the linear equation
\begin{equ}[e:SC]
d W_{\cL_\eps}(x,t) = \CL_\eps W_{\cL_\eps}(x,t)\,dt + \sqrt{Q_\eps}\,dW(x,t)\;,
\end{equ}
where
\begin{equ}
\cL_\eps = -1-\eps^{-2} P(\eps i\d_x)\;,
\end{equ}
$W$ is a standard cylindrical Wiener process on $\L^2([-L,L])$, and
the covariance operator $Q_\eps$ is given by the following definition.

\begin{definition}
\label{def:cov_op}
Let Assumption \ref{ass:gkeps} be true.
Define $q^\eps$ as the function such that $q_k^\eps$ are
its Fourier coefficients (cf. \eref{e:Dirk}).
Then define $Q_\eps$ as the rescaled convolution with  $q^\eps$,
\ie
\begin{equ}
\bigl(Q_\eps f\bigr)(x)
= {1 \over \eps} \int_{-L}^{L} f(y)\, q^\eps \Bigl({x-y \over \eps}\Bigr)\,dy\;.
\end{equ}
\end{definition}
Let us expand $W_{\cL_\eps}$ into a complex Fourier series.
Denote as usual by $e_k(x) = e^{ik\pi x/L}/\sqrt{2L}$
 the complex orthonormal Fourier basis on
$[-L,L]$. Define furthermore $P^\eps$ by
\begin{equ}
P^\eps(k) = {1\over \eps^2}P\Bigl({k\eps \pi\over L}\Bigr) + 1
\end{equ}
Since $Q_\eps$ commutes with $\cL_\eps$, we can write the stochastic convolution as
\begin{equs}
    W_{\cL_\eps}(x,t)
    &=  \sqrt{Q_\eps} \int_0^t e^{\cL_\eps(t-s)} dW(x,s) \\
    &= \sum_{k=-\infty}^\infty \sqrt{q_k^{\eps}}\, e_k(x) \int_0^t \exp\bigl({-P^\eps(k)\,(t-s)}\bigr)\,dw_k(s)\;,
\end{equs}
where the  $\{w_k\}_{k\in\Z}$ are complex standard Wiener processes
that are independent, except for the relation $w_{-k}=\overline{w_k}$.
We approximate $W_{\cL_\eps}(x,t)$
by expanding $P$ in a Taylor series up to order two around
its zeroes. We thus define the approximating polynomials $P_j^\eps$ by
\begin{equ}
P_j^\eps(k) = {P''(\zeta_j)\pi^2 \over 2L^2}\Bigl(k - {L \zeta_j\over \eps\pi}\Bigr)^2+1\;.
\end{equ}
With this notation, the approximation $\Phi(x,t)$ is defined by
\begin{equ}
 \label{e:defW0}
\Phi(x,t)
= 2\Re \sum_{j=1}^m \sqrt{\hat q(\zeta_j)} \sum_{k=-\infty}^\infty e_k(x) \int_0^t \exp\bigl({-P_j^\eps(k) (t-s)}\bigr)\,d\tilde w_{k,j}(s)\;,
\end{equ}
where the $\tilde w_{k,j}$'s are complex i.i.d.\ complex standard Wiener processes.
At this point, let us discuss a rewriting of $\Phi$ which makes the link with the notations
used in the rest of this article. We decompose ${L \zeta_j\over \eps\pi}$ into an integer
part and a fractional part, so we write it as
\begin{equ}
{L \zeta_j\over \eps\pi}
= \delta_j + k_j\;,\qquad \delta_j \in \bigl[{\textstyle{-\frac12}, \textstyle{\frac12}}\bigr]\;,
\qquad k_j = \Bigl[{L \zeta_j\over \eps\pi}\Bigr]\in\Z.
\end{equ}
As before $[z]$ denotes the nearest integer to $z\ge0$, with
the convention that $[\frac12]=1$. For $z<0$, we define $[z]=-[-z]$.
Extend for $m>1$ the definition
of the Hilbert space $\CH_a = \L^2([-L,L], \C^m)$ and the
definition of the projection
\begin{equs}
\pi_\eps: \CH_a &\mapsto \CH_u \\
A &\to 2 \Re \sum_{j=1}^m A_j(x) e^{{i \pi k_j \over L} x}\;.
\end{equs}
With this notation, we can write $\Phi$ as $\Phi(t) = \pi_\eps \Phi^a(t)$, where the
$j$-th component of $\Phi^a$ solves the equation
\begin{equ}[e:Phi]
d\Phi_j^a(t) = \Delta_j \Phi_j^a(t)\,dt + \sqrt{\hat q(\zeta_j)}\,\eta_j(t)\;.
\end{equ}
Here, the $\eta_j$'s are independent complex-valued space-time white noises and
the Laplacian-type operator $\Delta_j$ is given by
\begin{equ}
\Delta_j = -{P''(\zeta_j)\over 2} \Bigl(i\d_x + {\pi \delta_j\over L}\Bigr)^2\;.
\end{equ}
Now we can prove the following approximation result.
\begin{proposition} \label{prop:stoch_conv}
Let Assumptions \ref{ass:P}, \ref{ass:g} and \ref{ass:gkeps} hold and consider
$\Phi$ and $W_{\cL_\eps}$ as defined in \eref{e:SC} and \eref{e:Phi}.
Then for every $T>0$, $\kappa > 0$ and every $p \ge 1$,
there exists a constant $C$ and joint realisations of the noises $W$ and $\eta_i$ such
that
\begin{equ}
\EX \Big(\sup_{x \in [-L,L]} \sup_{t\in [0,T]} |\Phi(x,t) - W_{\cL_\eps}(x,t)|^p\Big) \le C \eps^{p/2-\kappa}\;.
\end{equ}
\end{proposition}
\begin{remark}
This result can not be generalised to dimensions higher than one, since the stochastic
convolution of the Laplace operator with space-time white noise is then not even in $\L^2$.
It the zeros of $P$ are degenerate, \ie $P$ behaves like $(k-\zeta_j)^{2d}$
for some $d\in \{2,3,\ldots\}$ then we would obtain an amplitude equation with
higher order differential operator, and we can proceed to higher dimension.
The other option would be to use fractional noise in space, which is more regular that
space-time white noise. Using the scaling invariance of fractional noise,
we would obtain fractional noise in the amplitude equation.
\end{remark}
\begin{proof}
It will be convenient for the remainder of the proof to distinguish between the positive roots $\zeta_j$ and
the negative roots $-\zeta_j$ of $P$, so we define $\zeta_{-j} = -\zeta_j$.
We start by writing $\Phi = \sum_{j=1}^m \bigl(\Phi^{(j)} + \Phi^{(-j)}\bigr)$ with
\begin{equ}[e:defPhij]
\Phi^{(j)}(x) = \left\{
\begin{array}{ll}
    \Phi^a_j(x) \, e^{{i \pi k_j \over L} x} & \text{for $j > 0$,} \\
    \overline{\Phi^a_j(x)}\, e^{- {i \pi k_j \over L} x} & \text{for $j < 0$.}
 \end{array}\right.
\end{equ}
For $r>0$ sufficiently small and $R$ as in \textbf{P5},
we decompose $\Z$ into several regions:
\begin{equs}
K_1^{(j)} &= \Bigl\{k \in \Z\,\Big|\, \Bigl|{k\eps\pi\over L} - \zeta_j \Bigr| < r\Bigr\}\;,
\quad K_1=K_1^{(0)} = \bigcup_{j=1}^m \bigl(K_1^{(j)}\cup K_1^{(-j)}\bigr)\;,\\
K_2 &= \Bigl\{k \in \Z\,\Big|\, \Bigl|{k\eps\pi\over L} \Bigr| < R\Bigr\}\;, \qquad
K_3 = \Z \setminus K_2\;.
\end{equs}
We suppose that $r>0$ is sufficiently small such that the $\{K_1^{(j)}\}_{j=\pm 1,\ldots,\pm m}$ are disjoint
and such that $0 \not \in K_1$.
The splitting into $K^2$ and $K^3$ is mainly for technical reasons.
We denote by $\Pi_1^{(j)}$, $\Pi_2$, etc.\ the corresponding orthogonal projection
operators in $\L^2([-L,L])$. We also define
\begin{equs}
\gamma_k= \gamma_k^{(0)}&= {1\over \eps^2}P\Bigl({k\eps \pi\over L}\Bigr) +1\;,\\
\gamma_k^{(j)} &= {P''(\zeta_j)\pi^2 \over 2L^2}\Bigl(k -\frac{L\zeta_j}{\eps\pi}\Bigr)^2 + 1\;
\quad\mathrm{for}\   j=\pm 1,\ldots,\pm m
\end{equs}
It is a straightforward calculation, using Taylor expansion
and  Assumption \ref{ass:P}, that there exist constants $c$ and $C$
independent of $\eps$ and $L$ such
that one has the following properties for $j=\pm 1,\ldots,\pm m$:
\minilab{e:propg}
\begin{equs}[2]
\bigl|\gamma_k - \gamma_k^{(j)}\bigr|
    &\le \frac{C \eps}{L^3} \Bigl|k - {\zeta_j L \over \pi\eps}\Bigr|^3\;,
        &\quad k&\in K_1^{(j)}\;,\label{e:propg1}\\
|\gamma_k^{(j)}|
    &\ge 1+\frac{c}{L^2} \Bigl|k - {\zeta_j L \over \pi\eps}\Bigr|^2\;,
        & k &\in K_1^{(j)}\;,\label{e:propg2}\\
|\gamma_k^{(j)}|
    &\ge {c\over \eps^2}\;,
            & k &\in K_2 \setminus K_1^{(j)}\;,\label{e:propg4}\\
|\gamma_k^{(j)}|
    &\ge {c k^2}/{L^2}\;,
            & k &\in K_3\;.\label{e:propg6}
\end{equs}
In view of the series expansion of Lemma \ref{lem:KLexp}, we also define
\begin{equ}[e:defank]
a_{n,k}^{(j)} = C \sqrt{ {1- (-1)^n  e^{-\gamma_k^{(j)} T} \over (\gamma_k^{(j)})^2 T^2 + \pi^2 n^2}}\;,
\end{equ}
where the constant $C$ depends only on $T$. We define $a_{n,k}$ in the same way with
$\gamma_k^{(j)}$ replaced by $\gamma_k$.
With these definitions at hand, we can
use Lemma \ref{lem:KLexp} to write $\Phi^{(j)}$ as
\begin{equ}
 \Phi^{(j)}(t,x) = \sqrt{\hat q(\zeta_j)} \sum_{k = -\infty}^\infty
\sum_{n \in \Z} a_{n,k}^{(j)} \xi_{n,k}^{(j)} e_{n,k}^{(j)}(x,t)\;,
\end{equ}
where we defined
\begin{equ}
e_{n,k}^{(j)}(x,t) = e_k(x) \bigl(e^{{i\pi n\over T}t} - e^{-\gamma_k^{(j)}t}\bigr)\;,
\end{equ}
and where the $\{\xi_{n,k}^{(j)}:\ n\in\Z\}$
are independent complex-valued Gaussian random variables.
Note that $e_{-n,-k}^{(-j)}(x,t) = \overline{e_{n,k}^{(j)}(x,t)}$, so that \eref{e:defPhij}
implies the relation
$\xi_{-n,-k}^{(-j)} = \overline{\xi_{n,k}^{(j)}}$.
The process $W_{\cL_\eps}(t,x)$ can be expanded in a similar way as
\begin{equ}[e:expWeps]
W_{\cL_\eps} (t,x)=
\sum_{k = -\infty}^\infty \sqrt{ q_k^\eps } \sum_{n \in\Z} a_{n,k} \xi_{n,k}e_{n,k}(x,t)\;,
\end{equ}
with
\begin{equ}
e_{n,k}(x,t) = e_k(x) \bigl(e^{{i\pi n\over T}t} - e^{-\gamma_k t}\bigr)\;,
\end{equ}
where $\{\xi_{n,k}:\ n\in\Z, k\in\Z\}$ are i.i.d\
standard complex-valued Gaussian random variables, with the exception
that $\xi_{-n,-k}=\overline{\xi_{n,k}}$. Note that this implies that $\xi_{0,0}$ is real-valued.
In order to be able to compare $W_{\cL_\eps}$ and $\Phi$, we now specify how we choose
the random variables $\xi_{n,k}$ to relate to the random variables $\xi_{n,k}^{(j)}$.
For $j=\pm 1,\ldots,\pm m$ we define $\xi_{n,k}^{(j)}:=\xi_{n,k}$ for all $k\in K_1^{(j)}$.
Note that this is consistent with the relations $\xi_{-n,-k}^{(-j)} = \overline{\xi_{n,k}^{(j)}}$
and $\xi_{-n,-k}=\overline{\xi_{n,k}}$, and with the fact that $K_1^{(-j)} = -K_1^{(j)}$.
We will see later in the proof that the definition of  $\xi_{n,k}^{(j)}$
for $k \not\in K_1^{(j)}$ does not really matter, so we choose them to be independent
of all the other variables, except for the relation $\xi_{-n,-k}^{(-j)} = \overline{\xi_{n,k}^{(j)}}$.
The the proof of the proposition is split into several steps.
First we bound the difference of ${1\over2}\Pi_1^{(j)}\Phi^{(j)}$ and $\Pi_1^{(j)}W_{\cL_\eps}$.
Then we show that all remaining terms  $(1-\Pi_1^{(j)})\Phi^{(j)}$
and $(1-\Pi_1^{(0)}) W_{\cL_\eps}$ are small.
\begin{step}
We first prove that for $j=\pm1,\ldots,\pm m$
\begin{equ}[e:bound1]
\EX \sup_{x \in [-L,L]} \sup_{t\in [0,T]} |\Pi_1^{(j)} \Phi^{(j)}(x,t) - \Pi_1^{(j)}W_{\cL_\eps} (x,t)|^p \le C \eps^{p/2-\kappa}\;.
\end{equ}
We thus want to apply \lem{lem:Dirk} to
\begin{equ}
I(t,x) := \sum_{k \in K_1^{(j)}} \sum_{n \in\Z} \xi_{n,k}  \bigl(
\sqrt{\hat q(\zeta_j)} a_{n,k}^{(j)} e_{n,k}^{(j)}(x,t) - \sqrt{q_k^\eps} a_{n,k} e_{n,k}(x,t)\bigr)
\end{equ}
Define
$$ f_{n,k}(x,t)
    = \sqrt{\hat q(\zeta_j)} a_{n,k}^{(j)} e_{n,k}^{(j)}(x,t)
        - \sqrt{q_k^\eps} a_{n,k} e_{n,k}(x,t).
$$
Note first that $\mathrm{Lip}(f_{n,k}) \le C(1 + |k| + |n| + |\gamma_k|)$ and similarly
for $\mathrm{Lip}(f_{n,k}^{(j)})$. Therefore, the uniform bounds on $\hat q$ and $q_k^\eps$,
together with the definition of $a_{n,\gamma}$
imply that there exists a constant $C$ such that $\mathrm{Lip}(f_{n,k})$ is bounded by $C(|k|+1)$
for all $k\in K_1^{j}$ and $n\in\N$, where the constant only depends on $T$.
Note that the Lipschitz constant is taken with respect to $x$ and $t$.
For $k\in K_1^{(j)}$ we have  $|k| \le C/\eps$,
and hence $\mathrm{Lip}(f_{n,k})\le C\eps^{-1}$.
Now  \lem{lem:Dirk} implies \eref{e:bound1}
if we can show that for every $\kappa > 0$ one has
\begin{equ}[e:sumdeltan]
 \sum_{k \in K_1^{(j)}} \sum_{n\in\Z} \|f_{k,n}\|_\infty^{2-\kappa} \le C_\kappa \eps^{1-\kappa}\;,
\end{equ}
where the $L^\infty$-norm is again taken with respect to $t$ and $x$.
To verify \eref{e:sumdeltan} we estimate $\|f_{k,n}\|_\infty$ by
\begin{equs}
\|f_{k,n}\|_\infty &\le |\sqrt{\hat q(\zeta_j)} - \sqrt{q_k^\eps}| |a_{n,k}| \|e_{n,k}\|_\infty +
|\sqrt{\hat q(\zeta_j)}| |a_{n,k}^{(j)}| \|e_{n,k}^{(j)} - e_{n,k}\|_\infty\\
&\quad + |\sqrt{\hat q(\zeta_j)}| |a_{n,k}^{(j)}-a_{n,k}| \|e_{n,k}\|_\infty\\
&\eqdef I_1(n,k) + I_2(n,k) + I_3(n,k)\;,
\end{equs}
and we bound the three terms separately.
First by assumption $\|\hat q\|_\infty \le C$.
Furthermore,
$a_{n,k} \le C/(1+|n|)$  and $\|e_{k,n}\|_\infty \le C$ for all $k\in K_1^{(j)}$ and $n\in\N$,
and analogous for the terms involving $j$.
Again by assumption
$ |\sqrt{\hat q(k_j)} - \sqrt{q_k^\eps}| \le C\eps$ for all $k \in K_1^{(j)}$,
so that $I_1(n,k)$ is bounded by
\begin{equ}[e:boundI1]
|I_1(n,k)| \le {C\eps \over 1+|n|}\;.
\end{equ}
And hence, $\sum_{k,n}|I_1(n,k)|^{2-\kappa}\le C\eps^{1-\kappa}$.
For every $t > 0$ and every $\gamma' > \gamma > 0$
\begin{equ}
|e^{-\gamma t}-e^{-\gamma' t}| \le Ct |\gamma - \gamma'| e^{-\gamma t}\;.
\end{equ}
Combining this with \eref{e:propg1}
one has $\|e_{n,k}^{(j)} - e_{n,k}\|_\infty \le C\eps |k-\frac{\zeta_j L}{\pi\eps}|$
for $k \in K_1^{(j)}$.
Using
$$\sum_{n=-\infty}^\infty(a_{n,k})^{2-\kappa}
\le C\sum_{n=-\infty}^\infty(\gamma_k+|n|)^{\kappa-2}
 \le C/(\gamma_k(1+\gamma_k)),$$
we derive $\sum_{n=0}^\infty I_2(n,k)^{2-\kappa} \le C\eps^{2-\kappa}.$
Which gives the claim.
Concerning $I_3$,
a straightforward estimate using \eref{e:propg1} shows that
\begin{equ}
|I_3(n,k)| \le C |a_{n,k}-a_{n,k}^{(j)}|
= C\eps  \frac{1+ \left|k-\frac{\zeta_j L}{\pi\eps}\right|}{\gamma_k+|n|}\;.
\end{equ}
Using  $\sum_{n=-\infty}^\infty(\gamma_k+|n|)^{\kappa-2} \le C/(\gamma_k(1+\gamma_k))$
we derive $\sum_{n=-\infty}^\infty I_3(n,k)^{2-\kappa} \le \frac{C}{\gamma_k}\eps^{2-\kappa}$,
where we can use \eref{e:propg2}.
Combining all three estimates,  bound \eref{e:sumdeltan} follows now easily.
\end{step}
\begin{step}
We now prove that
\begin{equ}[e:bound2]
\EX \sup_{x \in [-L,L]} \sup_{t\in [0,T]} |\Pi_3 \Phi^{(j)}(x,t)|^p \le C \eps^{p/2-\kappa}\;,
\end{equ}
and
\begin{equ}[e:bound3]
\EX \sup_{x \in [-L,L]} \sup_{t\in [0,T]} |\Pi_3 W_{\cL_\eps}(x,t)|^p \le C \eps^{p/2-\kappa}\;.
\end{equ}
Both bounds are obtained in the same way, so we only show how to prove \eref{e:bound3}.
Using the bound on $q_k^\eps$,
\eref{e:defank} and \eref{e:propg6} for $a_{n,k}$,
 and the definition of $e_{n,k}$, we readily obtain the bounds
\begin{equ}
    \|q_k^\eps a_{n,k}e_{n,k}\|_\infty \le {C \over k^2 + |n|}\;,
    \qquad
    \mathrm{Lip}(q_k^\eps a_{n,k}e_{n,k}) \le Ck\;.
\end{equ}
Now \eref{e:bound3} follows immediately from \lem{lem:Dirk}, noticing that
\begin{equ}
\sum_{n\in\Z}^\infty (k^2+|n|)^{-\delta}\le C |k|^{2-2\delta}\;,\quad\text{for $|k|\ge1$ and $\delta>1$.}
\end{equ}
Furthermore,
$K_3$ only contains elements $k$ larger than $C\eps^{-1}$.
\end{step}
\begin{step}
For $j=0,\ldots,m$ we denote by $\Pi_{21}^{(j)}$
the projector associated to the set $K_2\setminus K_1^{(j)}$.
We show that
\begin{equ}
\EX \sup_{x \in [-L,L]} \sup_{t\in [0,T]} |\Pi_{21}^{(0)} W_{\cL_\eps}(x,t)|^p \le C \eps^{p/2-\kappa}\;,
\end{equ}
and in a completely similar way we derive
\begin{equ}
\EX \sup_{x \in [-L,L]} \sup_{t\in [0,T]} |\Pi_{21}^{(j)} \Phi^{(j)}(x,t)|^p \le C \eps^{p/2-\kappa}\;.
\end{equ}
 By  \eref{e:defank} and \eref{e:propg4} we get
\begin{equ}
    \|q_k^\eps a_{n,k}e_{n,k}\|_\infty \le {C \over \epsilon^{-2} + |n|}\;,
    \qquad
    \mathrm{Lip}(q_k^\eps a_{n,k}e_{n,k}) \le C\eps^{-1}\;.
\end{equ}
The estimate follows then again from \lem{lem:Dirk}, noticing that $K_2-K_1$ contains
less than $\CO(\eps^{-1})$ elements.
\end{step}
Summing up the estimates from all the previous steps concludes the proof.
\end{proof}
%
%
%
\begin{appendix}
\section{Technical Estimates}
%
%
%
\subsection{Series expansion for stochastic convolutions}
\label{subsec:ser_exp}
This section provides technical results
on series expansion and their regularity
of stochastic convolutions,  which are
necessary for the proofs.
\begin{lemma}\label{lem:Dirk}
Let $\{\eta_k\}_{k\in I}$ be i.i.d.\
standard Gaussian random variables (real or complex) with $k\in I$ an arbitrary
countable index set.
Moreover let $\{f_k\}_{k\in I}\subset W^{1,\infty}(G,\C)$
where the domain $G\subset \R^d$  has sufficiently smooth boundary
(\eg piecewise $C^1$).
Suppose there is some $\delta\in(0,2)$ such that
$$
S_1^2=\sum_{k\in I} \|f_k\|^2_{L^\infty}<\infty
\quad\mathrm{and}\quad
S_2^2=\sum_{k\in I} \|f_k\|^{2-\delta}_{L^\infty}\mathrm{Lip}(f_k)^\delta<\infty
$$
Define $f(\zeta)=\sum_{k\in I}  \eta_k f_k(\zeta)$.
Then, with probability one, $f(\zeta)$ converges absolutely for any $\zeta\in G$
and,
for any $p>0$, there is a constant depending only on $p$, $\delta$, and $G$ such that
$$\EX \|f\|^p_{C^0(G)} \le C(S_1^p+S_2^p)\;.
$$
\end{lemma}
\begin{proof} From the assumptions we immediately derive that
$f(x)$ and $f(x)-f(y)$ are a centred Gaussian for any $x,y\in G$.
Moreover, the corresponding series converge
absolutely. Using that the $\eta_k$ are i.i.d.,
we obtain
\begin{equs}
\EX |f(x)-f(y)|^2
&= \sum_{k\in I}|f_k(x)-f_k(y)|^2\\
&\le \sum_{k\in I}\min\{2\|f_k\|_{L^\infty}^2, \mathrm{Lip}(f_k)^2|x-y|^2\}\\
&\le 2 \sum_{k\in I} \|f_k\|_{L^\infty}^{2-\delta}
 \mathrm{Lip}(f_k)^\delta  |x-y|^\delta \\
 &= 2S_2^2  |x-y|^\delta\;,
\label{e:lem:Dirka}
\end{equs}
where we used that $\min\{a,bx^2\} \le a^{1-\delta/2} b^{\delta/2} |x|^\delta$
for any $a,b\ge0$.
Furthermore,
\begin{equation}
\label{e:lem:Dirkb}
\EX|f(x)|^2 \le  \sum_{k\in I} \|f_k\|_{L^\infty}^2=S_1^2\;.
\end{equation}
Consider $p>1$ sufficiently large and $\alpha>0$ sufficiently small.
Using Sobolev embedding (cf. \cite[Theorem 7.57]{adams})
and the definition of the norm of the fractional Sobolev space
in \cite[Theorem 7.48]{adams}
we derive for $\alpha p>d$  that
\begin{equs}
\EX \|f\|^p_{C^0(G)}
&\le  C \EX \|f\|^p_{W^{\alpha,p}(G)}\\
&\le  C \EX \int_G\int_G \frac{|f(x)-f(y)|^p}{|x-y|^{d+\alpha p}} dx dy
    +C \EX \int_G |f(x)|^p dx \\
&\le  C  \int_G\int_G \frac{(\EX|f(x)-f(y)|^2)^{p/2}}{|x-y|^{d+\alpha p}} dx dy
    +C  \int_G (\EX|f(x)|^2)^{p/2} dx \;,
\end{equs}
where we used that $f(x)$ and $f(x)-f(y)$ are  Gaussian.
Note that the constants depend on $p$.
Using \eref{e:lem:Dirka} and \eref{e:lem:Dirkb},
we immediately see that
$$ \EX \|f\|^p_{C^0(G)} \le CS_1^p + CS_2^p$$
provided $\alpha\in(0,\delta/2)$.
Note finally that we  needed $p>d/\alpha$ to
have the Sobolev embedding available.
The case of $p\le d/\alpha$ follows easily using H\"older inequality.
\end{proof}
\begin{lemma}
\label{lem:KLexp}
Let $\gamma \in \R$ and let
\begin{equ}
a(t) = \int_0^t e^{-\gamma(t-s)}\,dw(s)\;,
\end{equ}
with $w$ a standard complex  Wiener process,
\ie $\EX w(t)w(s)=0$ and $\EX w(t)\overline{w(s)}=\min\{t,s\}$.
Then, for $t \in [0,T]$, $a(t)$ has the
following representation:
\begin{equ}[e:OUrepr]
a(t) = \sum_{n\in\Z} a_{n,\gamma}\xi_n (e^{\pi i n t \over T} - e^{-\gamma t})\;,
\end{equ}
where the $a_{n,\gamma}$
are given by the Fourier-coefficients of $\frac1{2\gamma}e^{-\gamma|t-s|}$ on $[-T,T]$
\begin{equ}
a_{n,\gamma}^2 = C {1 - (-1)^n e^{-\gamma T} \over \gamma^2 T^2 + \pi^2 n^2}
\end{equ}
with some constant $C$ depending only on the time $T$.
and the $\{\xi_n\}_{n\in\Z}$ are i.i.d.\ complex normal random variables,
\ie $\EX\xi_n^2=0$ and $\EX|\xi_n|^2 = 1$.
\end{lemma}
\begin{proof}
The stationary Ornstein--Uhlenbeck process
\begin{equ}
\tilde a(t) = \int_{-\infty}^t e^{-\gamma(t-s)}\,dw(s)
\end{equ}
has the correlation function:
\begin{equ}
\expect \overline{\tilde a(t)}\tilde a(s) = {e^{-\gamma|t-s|} \over 2\gamma}\;.
\end{equ}
Expanding $e^{-\gamma|z|}$
in Fourier series on $[-T,T]$ we obtain
$$\tilde{a}(t)= \sum_{n\in\Z}a_{n,\gamma}\xi_ne^{i\pi nt/T}\;,$$
for i.i.d.\ normal complex-valued Gaussian random variables $\xi_n$.
The claim now follows from the identity
$a(t) = \tilde a(t) - e^{-\gamma t}\tilde a(0)$.
\end{proof}
%
%
%
%
\subsection{A-priori estimate for the amplitude equation}
\label{sec:apeftae}
This section summarises and proves technical a-priori estimates
for an equation of the type \eqref{e:ampl2}. Most of them are obtained
by standard methods and the proofs will be omitted.
The main non-trivial result is Theorem \ref{thm:fourier}
about the concentration in Fourier space. We consider the equation
\begin{equation}
\partial_t A= \alpha \partial_x^2 A +i\beta\partial_x A+ \gamma A -  c
|A|^2A+\sigma \eta
\label{e:amp2}
\end{equation}
with periodic boundary conditions on $ [-L, L]$, where $\alpha$ and $c$ are positive
and $\sigma,\gamma,\beta \in\R$ and $\eta$ denotes space--time white noise.

Equation \eqref{e:ampl2} is of the form
\eqref{e:amp2} with $\alpha = 4, \, \beta = -8 \delta_\eps$, $\gamma = \nu - 4 \delta_\eps$ and
$c = 3$ with $|\delta_\eps |\leq \frac{\pi}{2 L}$.
Obviously,  the constants $\beta$ and $\gamma$ are $\eps$-dependent,
but uniformly bounded in $\eps>0$,
which is a straightforward modification of the
result presented.

Further, we denote by
$\cW$ the complex cylindrical Wiener process such that $\partial_t\cW=\eta$.
Define the stochastic convolution
\begin{equ}[e:defphi]
\varphi = \sigma \cW_{\alpha\partial_x^2 -1}
\qquad\mathrm{and}\qquad
B=A-\varphi.
\end{equ}
Then
\begin{equation}    \label{e:eqB}
    \partial_t B
    = \alpha \partial_x^2 B+ i\beta\partial_x (B+\varphi)
    + \gamma B +(\gamma+1)\varphi -  c |B+\varphi|^2(B+\varphi).
\end{equation}
Of course this equation is only formal, as $\varphi$ is not differentiable.
But in what follows, we can always use smooth approximations of $\varphi$
to justify the arguments. The mild formulation of (\ref{e:eqB}) is
\begin{equs}\label{e:mildB}
B(t) &= e^{\alpha\partial_x^2 t}A(0)
    +i\beta \int_0^t \partial_x e^{\alpha\partial_x^2 (t-s)}(B+\varphi)(s) ds\\
&\quad +\int_0^t e^{\alpha\partial_x^2 (t-s)}\Big(\gamma B(s)+(\gamma+1)\varphi(s)
                    -c  |B+\varphi|^2(B+\varphi)(s)\Big)ds\;.
\end{equs}
We will use the following Lemma, which fails to be true in higher dimensions for
complex space-time white noise $\eta$.
\begin{lemma}       \label{lem:APBphi}
For any choice of  $q\ge1$ and $T_0>0$ there are constants
such that
$$ \sup_{t \in [0,T_0]} \EX \|\varphi(t) \|_{\CC_a^0}^q \le C
\quad \mathrm{and} \quad   \EX  \sup_{t\in[0,T_0]} \|\varphi(t) \|_{\CC_a^0}^q \le C.$$
\end{lemma}
The results of the previous lemma  are obviously also true if we replace the $C^0$-norm  by an  $L^p$-norm.
The constant then depends also on $p$. The proof of this lemma is standard see \eg \cite{DB-Ha:PIM} or
\cite[Theorem 5.1.]{Bl-MP-Sc:01}.
Now we easily prove  the following result via standard energy-type estimates for $A-\varphi$.
\begin{proposition}
For any choice of $p\ge1$, $q\ge1$, and $T_0>0$ there are constants
such that
$$ \sup_{t\ge T_0} \EX \|A(t) \|_{L^p_a}^q \le C,$$
with constant independent of $A(0)$.
Moreover, for any choice of $c_0>0$, $p\ge1$, $q\ge1$, and $T_0>0$ there are
constants
such that if $\|A(0)\|_{L^p_a}^q \le c_0$ then
$$
 \sup_{t \in [0,T_0]} \EX \|A(t) \|_{L_a^p}^q \le C
\quad\mathrm{and}\quad
 \EX  \sup_{t\in[0,T_0]} \|A(t) \|_{L^p_a}^q \le C.
$$
\end{proposition}
Now we can easily verify the following result
using the mild formulation of solutions.
\begin{proposition}
\label{prop:apriori}
For any choice of $c_0>0$,  $q\ge1$, and $T_0>0$ there are constants
such that if $\EX \|A(0)\|_{\cC_a^0}^{3q} \le c_0$ then
$$
 \EX  \sup_{t \in[0,T_0]} \|A(t) \|_{\cC_a^0}^q \le C.
$$
\end{proposition}

Note that it is sufficient for Proposition \ref{prop:apriori}
to assume that $A(0)$ is admissible.

\begin{remark}
We need the condition on the $3q$th moment of the initial conditions to ensure that
$\EX  \sup_{t\in[0,T_0]} \| B |B|^2(t) \|_{L^p_a}^{q} \le C$.
\end{remark}
In the following we establish that a solution $A$ of \eqref{e:amp2} with admissible initial
conditions, in the sense of Definition \ref{def:admA},
stays concentrated in Fourier space in the $\cC^0$-topology for all times.

\begin{theorem}
\label{thm:fourier}
Let $A(t)$ be the solution of \eqref{e:amp2} and assume that the initial conditions are
admissible. Then for every $p \geq 1$ and $T_0 >0$ there exist positive constants $\kappa, \, C_0$
with $ \kappa \leq 1$ such that
$$
\EX \sup_{t \in[0,T_0]}\| \Pi^c_{\delta/ \epsilon}A(t) \|_{\cC_a^0}^p\le
C\eps^{p/2-\kappa}\;,
$$
where $\Pi^c_{\delta/ \epsilon}$ was defined in \eref{e:proj}.
\end{theorem}
\begin{proof} We start by establishing the fact that admissible initial conditions are
concentrated in Fourier space.
According to Definition \ref{def:admA} the initial conditions
admit the decomposition $ A(0) = W_0 + A_1$.
Consider first the Gaussian part $W_0$.
We can use the series expansion of Remark \ref{rem:noise}
together with Lemma \ref{lem:Dirk} to verify
$$
\EX \|\Pi_{\delta/ \eps}^c W_0 \|_{\cC^0_a}^p \leq C_p \eps^{p/2 - \kappa}.
$$
Let now $\{A^1_k \}_{k \in \mathbb{Z}}$ denote the Fourier coefficients of $A_1$. We
use the fact that $A_1$ is bounded in $\cH^1_a$ to deduce
\begin{equs}
 \|\Pi_{\delta/ \eps} A_1 \|_{\cC_a^0}^2 & \leq  \Bigl(
\sum_{|k| \geq \frac{\delta}{\eps}} |A^1_k| \Bigr)^2
             \leq \sum_{|k| \geq \frac{\delta}{\eps}} |k|^{-2}
                                   \sum_{k \in \mathbb{Z}} |k|^2 |A^1_k|^2
          \\ & \leq  C \eps^{1 - \kappa} \|A_1\|^2_{1}\;.
\end{equs}
From the above estimates we deduce that
$$
\EX \|\Pi_{\delta/ \eps}^c A(0) \|_{\cC_a^0}^p
\leq C \eps^{p/2 - \kappa}\;.
$$
Let us consider  \eref{e:mildB}.
First using the boundedness of the semigroup
$$
\EX \|\Pi_{\delta/ \eps}^c e^{\alpha t\partial_x^2}A(0) \|_{\cC_a^0}^p
\le C \EX \|\Pi_{\delta/ \eps}^c A(0) \|_{\cC_a^0}^p
\leq C \eps^{p/2 - \kappa}\;.
$$
Using the factorisation method (see \eg \cite[Theorem 5.1.]{Bl-MP-Sc:01}) we easily get
for the stochastic convolution $\phi$ defined in \eref{e:defphi} the bound
\begin{equation}
 \EX \,\Bigl\| \sup_{t \in[0,T_0]} \Pi^c_{\delta/ \epsilon} \varphi(t)\Bigr\|^p
\le C \Big( \sum_{|k|\ge \delta/\eps } |k|^{-2+2\kappa}\Big)^{p/2}\le
C\eps^{p/2-\kappa}.
\label{e:st_conv_est}
\end{equation}
To proceed, we use the stability of the semigroup
and the embedding of $\cH^\zeta$ into $\cC_a^0$ for $\zeta\in(\frac12,1)$.
Using this, it is elementary to show that
\begin{equ}
    \| \Pi^c_{\delta/\eps} e^{ t\alpha\partial_{x}^2} h\|_{\CC_a^0}
    \le C e^{-c t\eps^{-2}}t^{-\zeta/2} \| h\|\;,
\end{equ}
for every $h \in \CH_a$. Hence
\begin{equs}
\left\| \Pi^c_{\delta/\eps} \int_0^t e^{(t-s)\partial_x^2} h(s) ds \right\|_{\CC_a^0} &\le C
\int_0^t e^{-C s \eps^{-2}}  s^{-\alpha/2}ds
\sup_{s\in[0,T]}\|h(s)\|\\
&\le C \eps^{2-\zeta} \sup_{s\in[0,T]}\|h(s)\|.
\end{equs}
Moreover, for $h=\sum h_ke_k$ by a crude estimate
\begin{equs}
\left\| \Pi^c_{\delta/\eps}\partial_x  \int_0^t e^{(t-s)\alpha\partial_x^2} h(s) ds
\right\|_{\CC_a^0}
&\le \sum_{|k|\ge \delta/\eps} \int_0^t |k|e^{-c(t-s) k^2 }|h_k(s)|ds\\
&\le C  \int_0^t e^{-C s \eps^{-2}}  s^{-(1+\zeta)/2}ds
\sup_{s\in[0,t]}\|h(s)\| \\
&\le C \eps^{1-\zeta} \sup_{s\in[0,t]}\|h(s)\|.
\end{equs}
Using \eref{e:mildB}, Proposition \ref{prop:apriori}, and \eref{e:st_conv_est}
and choosing $\zeta>\frac12$ sufficiently small (\eg $\zeta=\frac12+\frac{\kappa}{p}$),
it is now straightforward to verify the assertion first for $B$ and hence for $A$.
\end{proof}
\end{appendix}
\bibliographystyle{Martin}
\markboth{\sc \refname}{\sc \refname}
\bibliography{ref-mo}
\end{document}